\documentclass[12pt]{iopart}
\usepackage[utf8]{inputenc}
\expandafter\let\csname equation*\endcsname\relax
\expandafter\let\csname endequation*\endcsname\relax
\usepackage{amsmath}
\usepackage{cite}
\usepackage{tikz}
\usepackage{import} %include files
\usepackage{hyperref} % label equation and general\
\usepackage{graphicx} %figures, tables, etc
\usepackage{relsize} %big symbols
\usepackage{caption} % gia pollaples eikones
\usepackage{braket}  % gia bra kai ket
\usepackage{commath} % gia absolute value 
 \usepackage{amssymb} % gia analog symbol
\usepackage{subcaption}
 \usepackage{cleveref}
 \usepackage{url}
 \usepackage{hyperref}
\def\si{\sigma}
\def\be{\begin{equation}}
\def\ee{\end{equation}}
\def\ln{\textrm{ln}}
\def\a{\alpha}

\begin{document}
\nocite{*}
\title[]{Multifractality in aperiodic quantum spin chains}

\author{Dimitrios Voliotis}

\address{Instituto de F\'{\i}sica de S\~{a}o Carlos, Universidade de S\~{a}o Paulo,C.P. 369, S\~{a}o Carlos, S\~{a}o Paulo 13560-970, Brazil.}
\ead{dimitrios.voliotis@usp.br}
\vspace{10pt}
%\begin{indented}
%\item[]August 2017
%\end{indented}

\begin{abstract}
Recently has been investigated that the ground-state wavefunction of the one dimensional quantum spin-1/2 chain models is multifractal in general with non-trivial fractal dimension. We are studying this phenomena for the quantum Ising chain with aperiodic perturbation. By performing a block real-space renormalization approach, we obtain the ground-state wave function and we extract the generalized multifractal dimension and the multifractal spectrum. For a spin chain with negative wandering exponent the multifractal quantities have the same behavior with the unperturbed chain while for a spin chain with a vanishing wandering exponent are dependent on the coupling ratio. Finally, for a spin chain with positive wandering exponent, the multifractal quantities present a different non-linear behavior.
\end{abstract}
%%%%%%%%%%%%%%%%%%%%%%%%%%%%%%%%%%%%%%%%%%%%%%%%%%%%
%%%%%%%%%%%%%%%%%%%%%%%%%%%%%%%%%%%%%%%%%%%%%%%%%%%%
\section{Introduction}\label{int}
Quantum spin chains are a convenient laboratory for the studies of novel and unique phases of matter and especially for the quantum phase transitions ~\cite{sach}. The quantum Ising chain in the presence of a transverse magnetic field is a prototypical model on the studies of quantum phase transitions, defined by the Hamiltonian
\begin{equation}\label{eq:hamilt_ising}
 H=-J\sum_{i}\sigma_{i}^{z}\sigma_{i+1}^{z}-h\sum_{i}\sigma_{i}^{x},
\end{equation}
where $\sigma^{x,z}$ are the usual Pauli matrices. With $J$, we denote the ferromagnetic interactions between next-neighbor spins and with $h$, the transverse magnetic field placed on each lattice site. The model undergoes a quantum phase transition from a ferromagnetic ($J>h$) to paramagnetic phase ($J<h$). The two phases are separated by a quantum critical point, located at $\prod_i h_i =\prod_i J_i$.

%Multifractality is a general notion, introduced to describe irregular fluctuations of physical systems. Since in problems of non-linear physics appears a variety of complicated fractal objects and strange sets, multifractality was introduced to simplify some of them \cite{procha,ak}. As an example of a fractal measure,  consider a physical quantity, i.e. the magnetization in problems of critical phenomena, which can be described by dividing it into $N$ number of pieces. The size of the $i$-th piece is $\ell_i$ and the event occurring upon, it's described by a number $M_i$. The corresponding number of typical value of the magnetization in the position $i$ is expected to scale with the size $\ell_i$, with a critical exponent $y$. We can define a whole set of exponents $y_q$, that the typical values vary with the index $q$ as \cite{kada1,kada} $$(M_i )^q \sim \ell_{i}^{y_q } , \quad q=1,2,3,...$$
%
%Hasley \emph{et al.} \cite{kada}, they emphasize to the case where $M_i$ has a meaning of probability that some event will occur upon the $i$-th piece. Also, the values of the exponents $y_q$, can be important for the understanding of the physical problem in general. 

For a quantum many-body system, an interesting questions arises whether a quantum wavefunction is multifractal and especially, the ground-state wave function. In the basis of $z$-components of each spin $\sigma_i$, of a system containing $N$ number of spins, $\ket{\vec{\si}}=\ket{\si_1 \cdots \si_N}$ with $\sigma_i =\pm1$, any wave function of spin-$1/2$ can be represented as 
\be\label{eq01}
\ket{\psi}=\sum_{\{\si_i \}}\psi_{\vec{\si}}\ket{\vec{\si}},
\ee
where we consider all the spin configurations of the Hilbert space of a size, $M=2^N$. The $M$ number of coefficients, they are following the normalization condition $1=\braket{\psi|\psi}=\sum\limits_{\{\si_i \} =1}^{M} \abs{\psi_{\vec{\si}}}^2$.

In this case, multifractality can occur to the coefficients $\psi_{\vec{\si}}$ of the wavefunction, and can scale non trivially with the Hilbert space dimension $M$ \cite{kada} as
\be\label{n1}
Y_q = \sum_{\{\si_i \}}\abs{\psi_{\vec{\si}}}^{2q} 
\underset{M\rightarrow\infty}{\sim} M^{-\tau_q},
\ee
where $Y_q$ is the so called Inverse Participation Ratio (IPR) \cite{an1,an2}. The exponent $\tau_q$ defines the \emph{generalized multifractal dimension exponent} $D_q$, by the relation $\tau_q =D_{q}(q-1)$.

Now consider as $S(q,M)$, the related R\'enyi entropy of the states of eq.~\eqref{eq01} of an $M\times M$ matrix, as
\be\label{eq03}
S(q,M)=-\frac{1}{q-1}\ln \Big( \sum_{\{\si_i \}}\abs {\psi_{\vec{\si}} }^2 \Big) .
\ee
Knowing the related R\'enyi entropy, we are able to determine the generalized multifractal dimension $D_q$, in the limit $M\rightarrow\infty$ \cite{mirlin1} by
\be\label{eq02}
D_{q}=\lim_{M\rightarrow\infty} \frac{S(q,M)}{\ln M} .
\ee
In general, $D_q$ has a non-linear dependence on $q$ in many different physical problems \cite{ref1,kada,ak,mirlin1,che,you,chr}. The physical quantity which preserves this behavior can be labeled as \emph{multifractal} by the fact that it can be characterized by some non-integer dimensions, known as fractal dimensions. By extracting the values of the fractal dimensions and in our case, the generalized fractal dimension $D_q$, we can define the mutlifractality which is a theory that it can describe and simplicity the physical problem. In quantum systems, especially for the disordered systems, such as the Anderson model, the wavefunctions shows highly nontrivial fluctuations \cite{eve}. These fluctuations can be precisely described by a multifractal analysis by extracting the exponents $D_q$. These kind of measurements has been recently applied  to the experimental study of disordered conductors \cite{exp1} and cold atoms \cite{exp2,exp3}.

Atas and Bogomolby \cite{atas1,atas2}, they proved by analytical and numerical studies that the ground-state wavefunction of a quantum spin chain models is multifractal. They considered chains that they can be mapped into a free-fermion model \cite{lieb}, extracting analytical results and they confirmed them by an exact diagonalization numerical technique. For the quantum Ising chain, the generalized multifractal dimension is equal to zero in the ferromagnetic phase, while is equal to one in the paramagnetic. On the other side, the scaling of the Shannon-R\'enyi entropy of eq.~\eqref{eq03}, has been studied extensively for the Shannon value $q=1$ as well as for the R\'enyi $q\geq2$, for several spin chains \cite{step1,step2} but without to emphasize to the existence of multifractal behavior in the system. Mirlin and Evers \cite{mirlin1}, they studied the multifractal properties for the critical fluctuations of various quantities for the Anderson transition. They found the generalized multifractal dimension $D_q$ of eq.~\eqref{eq02} to have different values in the localized and delocalized phase, while for a general case, the $D_q$ has a non-linear behavior with $q$. Finally, Monthus \cite{month0} using an analytical block real-space renormalization group scheme, she confirmed the basic features of the $D_q$ as well as she studied the Shannon-R\'enyi entropy for the pure and random quantum Ising chain.

A second quantity of multifractal formalism \cite{kada,mirlin1} that we are going to study here is the \emph{multifractal spectrum} or singularity spectrum $f(\a)$, defined by
\be
N_{M}(\a) \underset{M\rightarrow\infty}{\propto} M^{f(\a)} ,
\ee
where $N_{M}(\a)$ is the number of configurations $m$ having 
a weight $\abs{\psi_\sigma }^2 \propto M^{-\a}$.
Considering the saddle-point method, the IPR of eq.~\eqref{n1} for large $M$ can be written \cite{mirlin1,atas2,month0} as
\be
Y_q (M) \sim M^{-\tau(q)} \quad\textrm{and}\quad \tau(q)=q\a-f(\a),\quad \a=\tau'(\a),
\ee
where the miltifractal spectrum can be calculated via Legendre transform 
\be
f(\a)=q\a-\tau(q),\quad \a=\tau'(q) .
\ee

In this article, we are going to study the behavior of multifractal quantities, $D_q$ and $f(\a)$ for the quantum Ising model in a presence of an aperiodic perturbation.
We will consider different kind of aperiodicy, such as bounded, unbounded and the intermediate case. Our motivation is to study the evolution of different kind of aperiodic modulation on the multifractal quantities. Each aperiodic modulation lead the system to a different critical point and therefore is interesting to study how this is able to affect the multifractal quantities.
In practice, we are extending the block real-space renormalization procedure of Monthus \cite{month0,month1} for the aperiodic quantum Ising chain. The article is organized as follows. In section \ref{aper}, we discuss the properties of an aperiodic sequences generated through substitutions and the relevance/irrelevance criterion. In section ~\ref{block}, we are presenting the idea of the block real-space renormalization for the aperiodic quantum Ising chain. 
In section \ref{multi}, we extract the multifractal quantities for the Period-Doubling sequence and we compare them with the other aperiodic sequences of the table \ref{tabl1}. Finally, our conclusions are summarized in section \ref{end}. 
%
%%%%%%%%%%%%%%%%%%%%%%%%%%%%%%%%%%%%%%%%%%%%%%%%%%%%%
%
%%%%%%%%%%%%%%%%%%%%%%%%%%%%%%%%%%%%%%%%%%%%%%%%%%%%%
\section{Aperiodic sequences}\label{aper}
The context of aperiodicity is connected with the studies of the critical phenomena since it is related with several properties of the experimental area of quasicrystals \cite{quasi}. The aperiodic scheme, has the characteristic to introduce in to the system, different types of spatial heterogeneities with an example from mathematics the Penrose tilling. The
aperiodic modulation are generated by iterated application of substitution rules on letters $A$, $B$,... such as $A\rightarrow S(A)\rightarrow S(S(A))\rightarrow...$ and $B\rightarrow S(B)\rightarrow S(S(B))\rightarrow...$  
The properties of a given sequence are controlled by a substitution matrix $M$, defined by the corresponding rules and contain the number of letters $A,B,...$ in $S(A),S(B),...$ as 
\be
M=
\begin{pmatrix}
  n_{A}^{S(A)} & n_{A}^{S(B)} & ... \\
  n_{B}^{S(A)} & n_{B}^{S(B)} & ...\\
  \vdots & \vdots
 \end{pmatrix},
 \ee
where for example the matrix element $n_A^{S(A)}$ gives the number of letter $A$ in the substitution pattern $S(A)$.
As an example, the Period-Doubling sequence defined by the substitution rules, $S(A)\rightarrow AB$ and $S(B)\rightarrow AA$, has the following substitution matrix,
\be
M_{\textrm{Period-Doubling}}=
\begin{pmatrix}
  1 & 1 \\
  2 & 0
 \end{pmatrix}.
 \ee
In the table~\ref{tabl1}, we are presenting the three different types of aperiodic sequences that we will consider in this article.
\begin{table}[h!]
\begin{center}
\begin{tabular}{ |p{3cm}|p{3cm}|p{3cm}|  }
 \hline
 \multicolumn{3}{|c|}{Aperiodic Sequences} \\
 \hline
 Thue-Morse $(\omega<0)$ & Period-Doubling $(\omega=0)$ & Rudin-Shapiro $(\omega>0)$ \\
 \hline
  \multicolumn{3}{|c|}{Substitution Rules} \\
  \hline
  $A\rightarrow AB$, $B\rightarrow BA$ & $A\rightarrow AB$, $B\rightarrow AA$ & $A\rightarrow AB$, $B\rightarrow AC$, $C\rightarrow DB$, $D\rightarrow DC$ \\ 
 % \hline
  \hline
  \multicolumn{3}{|c|}{Iterations (n=4)} \\
  \hline
 $A$        & $A$        & $A$ \\
$AB$       & $AB$       & $AB$\\
$ABBA$     & $ABAA$     & $ABAC$\\
$ABBABAAB$ & $ABAAABAB$ & $ABACABDB$\\
.........  & .........  & .........\\
 \hline
\end{tabular}
\end{center}
\caption{Definition of the substitution rules of the aperiodic sequences: Theu-Morse, Period-Doubling and Rudin-Shapiro. In the last box is presented the first four iterations of the subsection rules.}
\label{tabl1}
\end{table}
With $\omega$, we denote the \emph{wandering exponent}, defined by the ratio of the largest with the second-largest eigenvalues of the  substitution matrix. From the substitution matrix $M$, several characteristics of the sequence can be extracted, by making use of the right eigenvectors $V_A$ of $M$,
$$ MV_A = \Lambda_A V_A ,$$
such as the length $L_n$ of the sequence, which is asymptotically proportional to the largest eigenvalue of $M$. The asymptotic density of the letters $A$ in the sequence, is given as a function of the corresponding eigenvectors,
$$\rho_{\infty}^{A}=\frac{V_{1}(1)}{\sum_i V_1 (i)}.$$
The exponent $\omega$, is the measure of the geometric fluctuations introduced into the system by the application of the substitution rules \cite{boo}. For negative $\omega$, the fluctuations are bounded and become negligible as the system size is grow by the iteration procedure. On the other side, for positive $\omega$, the fluctuations are unbounded and they grow randomnly with the iteration of the letter substitution rules. Finally, for vanishing $\omega$, the fluctuations grow logarithmicaly with the iteration procedure.

A similar argument of the Harris criterion \cite{harris} for the relevance of the disordered layered perturbations, introduced by Luck \cite{luck} for the relevance of the aperiodic perturbation. According to Luck's criterion, aperiodic modulation may be relevant, marginal or irrelevant, depending on the correlation length critical exponent $\nu$ of the unperturbed system and on the wandering exponent $\omega$ of the corresponding aperiodic sequence. 
The Luck's criterion prediction for the sequences of the table \ref{tabl1}, are summarized to, irrelevant for the Thue-Morse, marginal for the Period-Doubling and relevant critical behavior for the quantum Ising chain. These conclusions confirmed with various techniques, such as, free-fermions \cite{loic}, and renormalization procedure \cite{herm}.
%
%%%%%%%%%%%%%%%%%%%%%%%%%%%%%%%%%%%%%%%%%%%%%%%%%%%%%%%%%%%
\section{Block Real-Space Renormalization Group approach}\label{block}
The type of block real-space renormalization (BRG) procedure which we are interested about, introduced by Fernandez-Pacheco \cite{ferna-pach} for the critical behavior of the pure quantum Ising chain. This self-dual procedure, reproduces the critical point of the model as well as the correlation length exponent $\nu$ in agreement with the exact solution. The method applied to the study of the critical point of other models too, like Potts \cite{pot} and Ashkin-Teller \cite{ash}. Recently, the method extended to higher dimensions for the pure and the random quantum Ising chain by Migazaki \emph{et al.} \cite{nish1,nish2} and Monthus \cite{month1}. For the random Ising chain \cite{month1}, the method confirms the critical exponent $\psi=1/2$ of the Infinite-Disorder Fixed-Point (IDFP) \cite{fisher}. Finally, the method applied to the multifractal properties of the pure and random quantum Ising chains for the ground-state quantum wave function \cite{month0} as well as for the excited-states \cite{month2}. 

The main idea \cite{ferna-pach,month1}, is to replace each block of spin $(\sigma_{i-1},\sigma_{i})$ with a new renormalized one $(\sigma_{R(i)})$. One of the block spin's is considered as a fixed parameter $S_0 =\pm1$, i.e. the spin $\sigma_i^z$ on the $z$-basis. The corresponding one-spin Hamiltonian of the block is diagonalized, producing two eigenvalues and eigenstates. The ground-state state, is obtained by the low-state and is dependent on $S_0$. Then, the method proceed by dropping the high-energy states and keeping the low-energy states. The projector is defined on the coarse-grained system and  the effective couplings are obtained and found to be smaller that the previous. We are presenting the approach with more details in the following. 

Here, we are extending the idea of Monthus \cite{month1} for the quantum Ising chain in the presence of an aperiodic perturbation. Due to the block nature of the approach, we are considering aperiodic sequences with a ``conventional'' double
letter generation of the substitution rules, like  these of the table~\ref{tabl1}. In order, to apply analytically the BRG approach for the aperiodic chain, we apply an aperiodic modulation both, to the interaction coupling $J$ and to the transverse field $h$. 
\subsection{Renormalization into a blocks}
As a representative example, we apply the BRG into the Period-Doubling quantum Ising chain. For convenience, we consider the substitution rules of the sequence with numbers instead of letters, like $0\rightarrow01$ and $1\rightarrow00$. 
According to the substitution rules, the exchange couplings $J$ and the transverse fields $h$ of the quantum Ising chain of the Hamiltonian of eq. \eqref{eq:hamilt_ising} are considered in the following manner of the two blocks as
\begin{equation*}\label{rul}
\begin{split}
& (\textbf{\textrm{block-a}}) \rightarrow 01\\
& (\textbf{\textrm{block-b}}) \rightarrow 00.
\end{split}
\end{equation*}
The BRG procedure of ref. the  \cite{month1} in now applied to all the possible blocks of the chain. The renormalization is divided into:
\begin{itemize}
 \item{\underline{\textbf{Block-a}}}: Renormalization of the block $J_{0}J_{1}$.
 
The Hamiltonian term for this block (represented in Fig.~\ref{fig:lat})  is 
  \begin{figure}[h]
    \centering
        \begin{tikzpicture}

 \draw [line width=1pt] (-1,0)--(3,0);

 \draw[fill](-1,0)[color=blue] circle [radius=0.07];
 \draw[fill](1,0)[color=blue] circle [radius=0.07];
 \draw[fill](3,0)[color=blue] circle [radius=0.07];

 \node [below] at (-1,0) {$i-1$};
 \node [below] at (1,0) {$i$};
\node [below] at (3,0) {$i+1$};

% the h

 \draw [line width=1pt] [->] (-1,0) -- (-1,0.8); 

 \draw [line width=1pt] [->] (1,0) -- (1,0.8);
% \draw [line width=1pt] [->] (1,0) -- (1,0.8);

 \node [left] at (-1,0.8) {$h_0$};
 \node [left] at (1,0.8) {$h_1$};

%................................................
\node [below] at (-0,0) {$J_0$};
 \node [below] at (2,0) {$J_1$};

\end{tikzpicture}
        \caption{Lattice representation of the block-$a$.}
        \label{fig:lat}
    \end{figure}
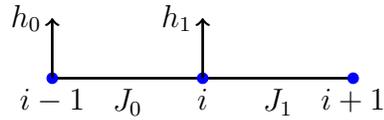
  \be\label{eq31}
  H_{a}=-J_{0}\si_{i-1}^{z}\si_{i}^{z}-J_{1}\si_{i}^{z}\si_{i+1}^{z}-h_{0}\si_{i-1}^{x}-h_{1}\si_{i}^{x}.
  \ee
The Hamiltonian is divided into the $intra$ and $extra$ term, $H_{a}=H_{intra}+H_{extra}$ as
\be\label{eq32}
H_{intra}=-J_{0}\si_{i-1}^{z}\si_{i}^{z}-h_{1}\si_{i}^x
\ee
and
\be\label{eq33}
H_{extra}=-J_{1}\si_{i}^{z}\si_{i+1}^{z}-h_{0}\si_{i-1}^x .
\ee
We consider the spin $\si_{i-1}^z$ as a fixed parameter into the two eigenvalues $S_{0}^{a}=\pm1$ and the $H_{intra}$ contained the interaction and transverse field in the spin $\si_i$. Therefore, we diagonalize the one-spin conventional Hamiltonian 
\be
\mathcal{H}_{a}^{(S_{0}^a)}=-J_{0}S_0^a \si_{i}^{z}-h_{1}\si_{i}^x
\ee
obtaining the eigenvalues
  \be
\lambda_{i}^{\pm}=\pm\sqrt{J_{0}^{2}+h_{1}^{2}}
 \ee
 and the eigenvectors 
 \be
\begin{split}
 & \ket{\lambda_{i}^{-}(S_{0}^{a})}=\sqrt{\frac{1+\frac{J_{0}S_{0}^{a}}{\sqrt{J_{0}^{2}+h_{1}^{2}}}}{2}}\ket{S_{i}=+}+\sqrt{\frac{1-\frac{J_{0}S_{0}^{a}}{\sqrt{J_{0}^{2}+h_{1}^{2}}}}{2}}\ket{S_{i}=-}\\
 & \ket{\lambda_{i}^{+}(S_{0}^{a})}=-\sqrt{\frac{1-\frac{J_{0}S_{0}^{a}}{\sqrt{J_{0}^{2}+h_{1}^{2}}}}{2}}\ket{S_{i}=+}+\sqrt{\frac{1+\frac{J_{0}S_{0}^{a}}{\sqrt{J_{0}^{2}+h_{1}^{2}}}}{2}}\ket{S_{i}=-}.
\end{split}.
\ee
The two corresponding ground-states for the block-$a$ are obtained by the tensor products
\be
\ket{GS(S_{0}^{a})} = \ket{S_{0}^{a}}\otimes \Big( \otimes_{i=1}^{N_a}\ket{\lambda_{i}^{-}(S_{0}^{a})}\Big),
\ee
where $N_a$ in the number of blocks-$a$ in the lattices.  
We proceed by keeping only the lowest-states, $\ket{\lambda_{i}^{-}(S_{0}^{a})}$ and dropping the highest-states $\ket{\lambda_{i}^{+}(S_{0}^{a})}$ in order to have access to the ground-state of the system by performing course-graining. For that reason, we define the projector onto the two ground-states
\be
P_{a}=\sum_{S_{i}^{a}=\pm1}\ket{GS(S_{0}^{a})} \bra{GS(S_{0}^{a})}.
\ee
The application of the projector in the $intra$ part of the Hamiltonian is producing the renormalized spin $\sigma_{R(i-1)}$ by  
\be
P_{a}\si_{i-1}^{z}P_{a}=\si_{R(i-1)}^z ,
\ee
\be
P_{a}\si_{i}^{z}P_{a}= \frac{J_{0}}{\sqrt{J_{0}^2 + h_{1}^2 }} \si_{R(i-1)}^z
\ee
\be
P_{a}\si_{i-1}^x P_{a} = \frac{h_{1}}{\sqrt{J_{0}^2 + h_{1}^2 }} \si_{R(i-1)}^x .
\ee
Then, the renormalized Hamiltonian is obtained by projection into the remaining part of the Hamiltonian for the eq. \eqref{eq31}, \eqref{eq32} and \eqref{eq33}
\be
H_{a}^{R}=P_{a}(H_a -H_{intra})P_{a},
\ee
giving the renormalized Hamiltonian for the block-a:
\be
H_{a}^{R}=-J_{a}\si_{R(i-1)}^{z}\si_{R(i+1)}^{z}-h_{a}\si_{R(i)}^{x},
\ee
with the renormalized interaction coupling 
\be\label{eq:renJa}
J_{a}=\frac{J_{0}J_{1}}{\sqrt{J_{0}^{2}+h_{1}^2}}
\ee
and the renormalized transverse field
\be\label{eq:renha}
h_{a}=\frac{h_{0}h_{1}}{\sqrt{J_{0}^{2}+h_{1}^2}}.
\ee
\item{\textbf{\underline{Block-b}}}: Renormalization of a block $J_{0}J_{0}$

Repeating the procedure for the block-b, we consider the block Hamiltonian $H_b=H_{\textrm{intra}}+H_{\textrm{extra}}$ as 
\be
H_{intra}=-J_{0}\si_{i-1}^{z}\si_{i}^{z}-h_{0}\si_{i}^x
\ee
and
\be
H_{extra}=-J_{0}\si_{i}^{z}\si_{i+1}^{z}-h_{0}\si_{i-1}^x .
\ee
Therefore, we obtain the one-spin conventional Hamiltonian 
\be
\mathcal{H}^{(S_{0}^{b})}=-J_{0}S_0^b \si_{i}^{z}-h_{0}\si_{i}^x ,
\ee
extracting the eigenvalues
  \be
\lambda_{i}^{\pm}=\pm\sqrt{J_{0}^{2}+h_{0}^{2}}
 \ee
 and the eigenvectors 
 \be
\begin{split}
 & \ket{\lambda_{i}^{-}(S_{0}^{b})}=\sqrt{\frac{1+\frac{S_{0}^{b}J_{0}}{\sqrt{J_{0}^{2}+h_{0}^{2}}}}{2}}\ket{S_{i}=+}+\sqrt{\frac{1-\frac{S_{0}^{b}J_{1}}{\sqrt{J_{0}^{2}+h_{0}^{2}}}}{2}}\ket{S_{i}=-}\\
 & \ket{\lambda_{i}^{+}(S_{0}^{b})}=-\sqrt{\frac{1-\frac{S_{0}^{b}J_{0}}{\sqrt{J_{0}^{2}+h_{0}^{2}}}}{2}}\ket{S_{i}=+}+\sqrt{\frac{1+\frac{S_{0}^{b}J_{1}}{\sqrt{J_{0}^{2}+h_{0}^{2}}}}{2}}\ket{S_{i}=-}.
\end{split}.
\ee
For this block of spins by repeating the previous procedure on the ground-state projection on the lowest-lying states of the $\ket{\lambda_{i}^{-}(S_{0}^{b})}$, we obtain the renormalized couplings
\be
J_{b}=\frac{J_{0}^2}{\sqrt{J_{0}^{2}+h_{0}^2}} ,
\ee
and the renormalized transverse field
\be
h_{b}=\frac{h_0^2}{\sqrt{J_{0}^{2}+h_{0}^2}}.
\ee
\end{itemize}
\subsection{Renormalization procedure}
Our idea is to renormalize the lattice in a two different ways, instead of the clean case, where each block is renormalized in one way \cite{month1}. For the Thue-Morse (more details in the ~\ref{ap11}) and Period-Doubling sequences that defined by two substitution rules, we define two types of renormalized blocks. For the Rudin-Shapiro sequence  with four substitution rules (more details in the ~\ref{ap21}), we define four types of renormalized blocks correspondingly.

We present here the renormalization procedure for the Period-Doubling sequence. The number $0$, occupy a part of the chain which is given by the asymptotic density $\rho_{\infty}^{0}=2/3$. Then, the number $0$, will define $N_{a}=2N/3$ number of blocks-$a$. Correspondingly, the block-$b$, will follow the asymptotic density of the number $1$ and will generate a $N_{b}=N/3$ number of blocks.

\begin{figure}[ht]
%    \centering
        \begin{tikzpicture}

 \draw [line width=1pt] (-12,0)--(4,0);

  \draw [fill] (-12,0)[color=blue] circle [radius=0.07];
 \draw [fill] (-11,0)[color=blue] circle [radius=0.07];
  \draw [fill] (-10,0)[color=blue] circle [radius=0.07];
    \draw [fill] (-9,0)[color=blue] circle [radius=0.07];
 \draw [fill] (-8,0)[color=blue] circle [radius=0.07];
    \draw [fill] (-7,0)[color=blue] circle [radius=0.07];
  \draw [fill] (-6,0)[color=blue] circle [radius=0.07];
    \draw [fill] (-5,0)[color=blue] circle [radius=0.07];
    \draw [fill] (-4,0)[color=blue] circle [radius=0.07];
    \draw [fill] (-3,0)[color=blue] circle [radius=0.07];
     \draw [fill] (-2,0)[color=blue] circle [radius=0.07];
      \draw [fill] (-1,0)[color=blue] circle [radius=0.07];
      
  \draw [fill] (0,0)[color=blue] circle [radius=0.07] ;
   \draw [fill] (1,0)[color=blue] circle [radius=0.07];
    \draw [fill] (2,0)[color=blue] circle [radius=0.07];
     \draw [fill] (3,0)[color=blue] circle [radius=0.07];
      \draw [fill] (4,0)[color=blue] circle [radius=0.07];

% the h
\draw [line width=1pt] [->] (-12,0) -- (-12,0.8);
 \draw [line width=1pt] [->] (-11,0) -- (-11,0.8);
\draw [line width=1pt] [->] (-10,0) -- (-10,0.8);
 \draw [line width=1pt] [->] (-9,0) -- (-9,0.8);
 \draw [line width=1pt] [->] (-8,0) -- (-8,0.8);
 \draw [line width=1pt] [->] (-7,0) -- (-7,0.8);
 \draw [line width=1pt] [->] (-6,0) -- (-6,0.8); 
 \draw [line width=1pt] [->] (-5,0) -- (-5,0.8);
 \draw [line width=1pt] [->] (-4,0) -- (-4,0.8);
 \draw [line width=1pt] [->] (-3,0) -- (-3,0.8);
 \draw [line width=1pt] [->] (-2,0) -- (-2,0.8);
 \draw [line width=1pt] [->] (-1,0) -- (-1,0.8); 

 \draw [line width=1pt] [->] (0,0) -- (0,0.8);
 \draw [line width=1pt] [->] (1,0) -- (1,0.8);
 \draw [line width=1pt] [->] (2,0) -- (2,0.8);
 \draw [line width=1pt] [->] (3,0) -- (3,0.8);
 \draw [line width=1pt] [->] (4,0) -- (4,0.8);  
 
 \node [left] at (-12,0.8) {$h_0$};
 \node [left] at (-11,0.8) {$h_1$};
 \node [left] at (-10,0.8) {$h_0$};
 \node [left] at (-9,0.8) {$h_0$};
 \node [left] at (-8,0.8) {$h_0$};
 \node [left] at (-7,0.8) {$h_1$};
 \node [left] at (-6,0.8) {$h_0$};
 \node [left] at (-5,0.8) {$h_1$};
 \node [left] at (-4,0.8) {$h_0$};
 \node [left] at (-3,0.8) {$h_1$};
 \node [left] at (-2,0.8) {$h_0$};
 \node [left] at (-1,0.8) {$h_0$};
 \node [left] at (0,0.8) {$h_0$};
 \node [left] at (1,0.8) {$h_1$};
 \node [left] at (2,0.8) {$h_0$};
 \node [left] at (3,0.8) {$h_0$};
 \node [left] at (4,0.8) {$h_q$};

%................................................
\node [below] at (-11.5,0) {$J_0$};
 \node [below] at (-10.5,0) {$J_1$};
 \node [below] at (-9.5,0) {$J_0$};
 \node [below] at (-8.5,0) {$J_0$};
 \node [below] at (-7.5,0) {$J_0$};
 \node [below] at (-6.5,0) {$J_1$};
 \node [below] at (-5.5,0) {$J_0$};
 \node [below] at (-4.5,0) {$J_1$};
 \node [below] at (-3.5,0) {$J_0$};
 \node [below] at (-2.5,0) {$J_1$};
 \node [below] at (-1.5,0) {$J_0$};
 \node [below] at (-0.5,0) {$J_0$};
 \node [below] at (0.5,0) {$J_0$};
 \node [below] at (1.5,0) {$J_1$};
 \node [below] at (2.5,0) {$J_0$};
 \node [below] at (3.5,0) {$J_0$};

 %....................................................
 \fill[fill=yellow]
(-11,-1) node[fill=red!10,draw,double,rounded corners] {a-block};
  \fill[fill=yellow]
(-9,-1) node[fill=red!10,draw,double,rounded corners] {b-block};
 \fill[fill=yellow]
(-7,-1) node[fill=red!10,draw,double,rounded corners] {a-block};
 \fill[fill=yellow]
(-5,-1) node[fill=red!10,draw,double,rounded corners] {a-block};
 \fill[fill=yellow]
(-3,-1) node[fill=red!10,draw,double,rounded corners] {a-block};
 \fill[fill=yellow]
(-1,-1) node[fill=red!10,draw,double,rounded corners] {b-block};
 \fill[fill=yellow]
(1,-1) node[fill=red!10,draw,double,rounded corners] {a-block};
 \fill[fill=yellow]
(3,-1) node[fill=red!10,draw,double,rounded corners] {b-block};

\draw[ultra thick, ->][color=green] (-4,-1.8) arc (-5:40:-1);
 \node [left] at (-4,-2) {1st RG-step};
%...........................................................
%
%..............................................................
\draw [line width=1pt] (-8,-4)--(0,-4);

  \draw [fill] (-8,-4)[color=blue] circle [radius=0.07];
    \draw [fill] (-7,-4)[color=blue] circle [radius=0.07];
    \draw [fill] (-6,-4)[color=blue] circle [radius=0.07];
    \draw [fill] (-5,-4)[color=blue] circle [radius=0.07];
     \draw [fill] (-4,-4)[color=blue] circle [radius=0.07];
      \draw [fill] (-3,-4)[color=blue] circle [radius=0.07];
     \draw [fill] (-2,-4)[color=blue] circle [radius=0.07] ;
   \draw [fill] (-1,-4)[color=blue] circle [radius=0.07];
    \draw [fill] (0,-4)[color=blue] circle [radius=0.07];

 \node [below] at (-7.5,-4) {$J_a$};
 \node [below] at (-6.5,-4) {$J_b$};
 \node [below] at (-5.5,-4) {$J_a$};
 \node [below] at (-4.5,-4) {$J_a$};
 \node [below] at (-3.5,-4) {$J_a$};
 \node [below] at (-2.5,-4) {$J_b$};
 \node [below] at (-1.5,-4) {$J_a$};
 \node [below] at (-0.5,-4) {$J_b$};
% h
\draw [line width=1pt] [->] (-8,-4) -- (-8,-3.2); 
 \draw [line width=1pt] [->] (-7,-4) -- (-7,-3.2);
 \draw [line width=1pt] [->] (-6,-4) -- (-6,-3.2);
 \draw [line width=1pt] [->] (-5,-4) -- (-5,-3.2);
 \draw [line width=1pt] [->] (-4,-4) -- (-4,-3.2);
 \draw [line width=1pt] [->] (-3,-4) -- (-3,-3.2); 
 \draw [line width=1pt] [->] (-2,-4) -- (-2,-3.2);
 \draw [line width=1pt] [->] (-1,-4) -- (-1,-3.2);
 \draw [line width=1pt] [->] (0,-4) -- (0,-3.2);

\node [left] at (-8,-3.2) {$h_a$};
 \node [left] at (-7,-3.2) {$h_b$};
 \node [left] at (-6,-3.2) {$h_a$};
 \node [left] at (-5,-3.2) {$h_a$};
 \node [left] at (-4,-3.2) {$h_a$};
 \node [left] at (-3,-3.2) {$h_a$};
 \node [left] at (-2,-3.2) {$h_a$};
 \node [left] at (-1,-3.2) {$h_b$};
 \node [left] at (0,-3.2) {$h_q$};

 \fill[fill=yellow]
(-7,-5) node[fill=red!10,draw,double,rounded corners] {$a'$-block};
 \fill[fill=yellow]
(-5,-5) node[fill=red!10,draw,double,rounded corners] {$b'$-block};
 \fill[fill=yellow]
(-3,-5) node[fill=red!10,draw,double,rounded corners] {$a'$-block};
 \fill[fill=yellow]
(-1,-5) node[fill=red!10,draw,double,rounded corners] {$a'$-block};

\draw[ultra thick, ->][color=green] (-3.8,-5.8) arc (-5:40:-1);
 \node [left] at (-4,-6) {2nd RG-step};
%......................................................................................................................................
%
%....................................................................................................................................
\draw [line width=1pt] (-6,-8)--(-2,-8);

 \draw [fill] (-6,-8)[color=blue] circle [radius=0.07];
     \draw [fill] (-5,-8)[color=blue] circle [radius=0.07];
      \draw [fill] (-4,-8)[color=blue] circle [radius=0.07];
     \draw [fill] (-3,-8)[color=blue] circle [radius=0.07] ;
\draw [fill] (-2,-8)[color=blue] circle [radius=0.07] ;

\node [below] at (-5.5,-8) {$J_{a'}$};
 \node [below] at (-4.5,-8) {$J_{b'}$};
 \node [below] at (-3.5,-8) {$J_{a'}$};
 \node [below] at (-2.5,-8) {$J_{a'}$};

\draw [line width=1pt] [->] (-6,-8) -- (-6,-7.2);
 \draw [line width=1pt] [->] (-5,-8) -- (-5,-7.2);
 \draw [line width=1pt] [->] (-4,-8) -- (-4,-7.2); 
 \draw [line width=1pt] [->] (-3,-8) -- (-3,-7.2);
 \draw [line width=1pt] [->] (-2,-8) -- (-2,-7.2);

\node [left] at (-6,-7.2) {$h_{a'}$};
 \node [left] at (-5,-7.2) {$h_{b'}$};
 \node [left] at (-4,-7.2) {$h_{a'}$};
 \node [left] at (-3,-7.2) {$h_{a'}$};
 \node [left] at (-2,-7.2) {$h_{q}$};

\fill[fill=yellow]
(-5,-9) node[fill=red!10,draw,double,rounded corners] {$a''$-block};
 \fill[fill=yellow]
(-3,-9) node[fill=red!10,draw,double,rounded corners] {$b''$-block};

\end{tikzpicture} 
        \caption{Block RG procedure for the Period-Doubling sequence for a lattice of $N=16$ spins. The first spin chain correspond to the initial 1-d lattice, the
        second chain the first RG iteration with couplings $J_a$, $J_b$ and transverse fields $h_a$, $h_b$. Finally, the last chain correspond to the third RG iteration. The field $h_q$ don't participate in the renormalization procedure.}
        \label{fig:rg1}
    \end{figure}
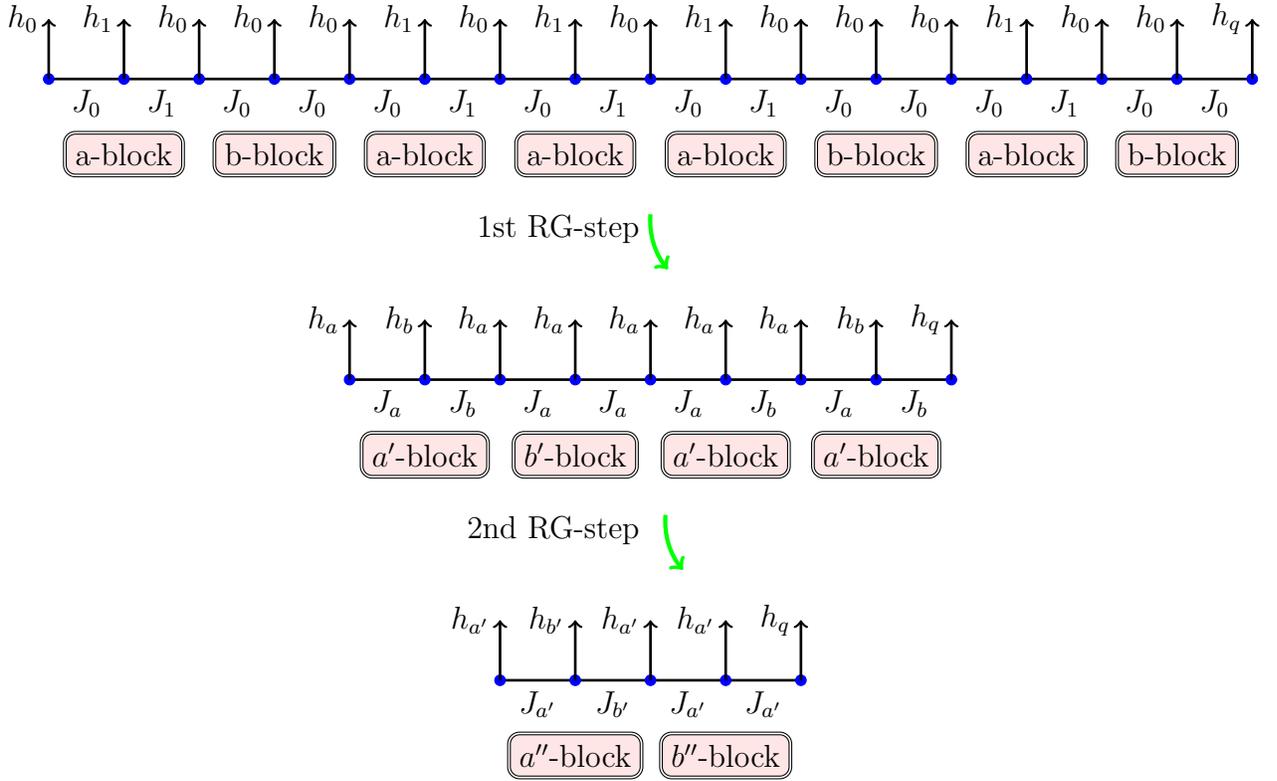

In Fig.~\ref{fig:rg1}, we present the BRG procedure for an example of $N=16$ lattice sites. The chain in each RG iteration is divided into $N/2$. The lattice under the BRG evolution, in the first iteration is renormalized into the new couplings of $J_a$ and $J_b$. The new transverse fields
are the $h_a$ and $h_b$. The second iteration, creates the couplings $J_{a'}$, $J_{b'}$ and the fields $h_{a'}$, $h_{b'}$ and so on. We observe that
our choice into the block renormalization leaves the lattice invariant in each iteration procedure. Each time, the aperiodic chain has the same form with the 
initial case but with new renormalized couplings and fields.
In particular, for the block-a, the RG proceed as: $N_{a}\rightarrow N_{a'}/2\rightarrow N_{a''}/4 \cdot\cdot\cdot N_{a^k}/2^k$, for $k$-number of iterations. 
The same applies for the block-b.

%The chain is renormalized as 

%$$
%N  \rightarrow \frac{N_{a}}{2}+\frac{N_b}{2}\rightarrow \frac{N_{a'}}{4}+\frac{N_{b'}}{4}  \rightarrow \frac{N_{a''}}{8}+\frac{N_{b''}}{8} \rightarrow \cdots
%$$

\subsection{The critical point}\label{cri_poi}
In order to extract the quantum critical point of the system, we consider the average of the variance, $\ln h-\ln J$ from the factor $\Delta$ \cite{nish2} as  
\be\label{eq:average}
\Delta=\frac{1}{N}\sum_{i=1}^{N}\ln\frac{h_i}{J_i}.
\ee
In our aperiodic chain the renormalized quantity of eq.\eqref{eq:average} is written as a sum of the renormalized blocks
\begin{equation} \label{eq1}
\begin{split}
\tilde{\Delta} &  =\Delta_{a}+\Delta_{b} \\
 & = N_{a}\ln\Big(\frac{h_a}{J_a}\Big) + N_{b}\ln\Big(\frac{h_b}{J_b}\Big)\\
 & = \ln\Big(\frac{h_{0} h_1 }{J_{0}J_{1}}\Big)^{N_a}+\ln\Big(\frac{ h_{0}^2 }{J_{0}^2} \Big)^{N_b},
\end{split}
\end{equation}
where $N_a$ and $N_b$ are the asymptotic densities of the letters $0$ and $1$ of the Period-Doubling sequence (where $\rho_{\infty}^{0}=N_{a}=2/3$ and $\rho_{\infty}^{1}=N_{b}=1/3)$
and therefore asymptotic densities of the two blocks.

According to the Pfeuti \cite{pfe} exact solution for the quantum Ising chain, on the critical point, the interaction couplings are related with the magnetics
fields with the condition $\prod_{i}J_{i}=\prod_{i}h_{i}$. Then, the expression inside the logarithm on eq. \eqref{eq1} has to be equal to unity. On the critical point, the two transverse field $h_0$ and $h_1$ related to the two interaction coupling $J_0$ and $J_1$ by the relation
\be\label{eq:crp0}
h_{0}^2 h_{1}=J_{0}^{2}J_{1}.
\ee
%
%If we consider a uniform transverse field, $h_{0}%=h_{1}=h$, them the above relation for the first RG %iteration is 
%
%\be\label{eq:crp}
%h_{crit}=J_{0}^{2/3}J_{1}^{1/3} .
%\ee
%
%
%%%%%%%%%%%%%%%%%%%%%%%%%%%%%%%%%%%%%%%%%%%%%%%%%%%%%%%%
\section{Multifractality for an aperiodic quantun spin chain}\label{multi}
\subsection{BRG rules for the IPR}\label{mul1}
According to our previous discussion about the BRG evolution in the aperiodic system, the IPR will follow the same procedure. In particular, the IPR is divided into the $Y_{q}^{a}$ for the block-a and $Y_{q}^{b}$ for the block-b. Following the procedure for the clean chain \cite{month0}, the IPR of the eq.~\eqref{n1} under the BRG approach is given as
\begin{equation} \label{eq2}
\begin{split}
Y_{q}(N) &  = \sum_{S_{1}=\pm1} \sum_{S_{2}=\pm1}... \sum_{S_{N}=\pm1} \abs{\psi(S_{1},S_{2},...,S_{N})}^{2q}\\
         & =  \Bigg( \sum_{S_{1}=\pm1}... \sum_{S_{N_{a}}=\pm1} \bigg[\abs{\psi_{a}^{R}}^{2q}\prod_{i=1}^{N_{a}/2}\tilde{y}_{q}^{a}(i)  \bigg] \Bigg)
        \Bigg( \sum_{S_{1}=\pm1}... \sum_{S_{N_{b}}=\pm1} \bigg[\abs{\psi_{b}^{R}}^{2q}\prod_{i=1}^{N_{b}/2}\tilde{y}_{q}^{b}(i)  \bigg] \Bigg) . \\
\end{split}
\end{equation}
The $\tilde{y}_{q}^{a,b}(i-1;i)$ represents the IPR of the wave-functions $\lambda_{i}^{(-)a,b}(S_{0}^{a,b})$ of the two blocks, given by
%|
\be\label{eq3}
\tilde{y}_{q}^{a,b}(i-1;i)\equiv \sum_{S_{i}^{a,b}=\pm1}\abs{\braket{S_{i}^{a,b}|\lambda_{i}^{(-)a,b}(S_{0}^{a,b})}}^{2q}.
\ee
For each block, the IPR for the wave-function is written as
\be
\tilde{y}_{q}^{a}(i-1;i)=\Bigg[\frac{1+\frac{J_0}{\sqrt{h_{1}^{2}+J_{0}^2}}}{2}\Bigg]^{q} +
\Bigg[\frac{1-\frac{J_0}{\sqrt{h_{1}^{2}+J_{0}^2}}}{2}\Bigg]^{q}
\ee
and 
\be
\tilde{y}_{q}^{b}(i-1;i)=\Bigg[\frac{1+\frac{J_0}{\sqrt{h_{0}^{2}+J_{0}^2}}}{2}\Bigg]^{q} +
\Bigg[\frac{1-\frac{J_0}{\sqrt{h_{0}^{2}+J_{0}^2}}}{2}\Bigg]^{q}.
\ee
The above expressions are independent of the two spins $S_{0}^a$ and $S_0^b$ and by introducing the ratios $K_{a}=J_{0}/h_{1}$ and $K_{b}=J_{0}/h_{0}$, we
define the auxiliary functions
\be\label{au1}
y_{q}^{a}=\Bigg[\frac{1+\frac{K_a}{\sqrt{1+K_{a}^2}}}{2}\Bigg]^{q} +
\Bigg[\frac{1-\frac{K_a}{\sqrt{1+K_{a}^2}}}{2}\Bigg]^{q},
\ee
\be\label{au2}
y_{q}^{b}=\Bigg[\frac{1+\frac{K_b}{\sqrt{1+K_{b}^2}}}{2}\Bigg]^{q} +
\Bigg[\frac{1-\frac{K_b}{\sqrt{1+K_{b}^2}}}{2}\Bigg]^{q}.
\ee
Then, the BRG rules for IPR of eq.\eqref{eq2}, can be summarized as
\be\label{eq34}
Y_{q}(N)=Y_{q}^{a}(N_{a})Y_{q}^{b}(N_{b})
\ee
where
\be\label{eq35}
Y_{q}(N_{a,b})=\bigg(\prod_{i=1}^{N_{a,b}/2}y_{q}(K_{a,b})\bigg)Y_{q}^{R}(\frac{N_{a,b}}{2})
\ee
and 
\be\label{eq36}
Y_{q}^{R}(\frac{N_{a,b}}{2})=\sum_{S_{1}}\sum_{S_{2}}\cdot\cdot\cdot\sum_{S_{N_{a,b}}}\abs{\psi^{R}(S_{1},S_{2},...,S_{N_{a,b}})}^{2q}.
\ee
The renormalization procedure of IPR of eq.~\eqref{eq34}, given as function of the IPR of the two blocks of eq.~\eqref{eq35} and \eqref{eq36}, is controlled by the ratio couplings $K_a$ and $K_b$ as well as by the numbers $N_a$ and $N_b$ which they are invariant in each RG iteration.
\subsection{BRG rules for the Shannon-R\'enyi entropy}\label{mul2}
According to our analysis so far, the Shannon-R\'enyi entropy of eq.~\eqref{eq03} and by considering the expression of the IPR of eq.~\eqref{n1}, the Shannon-R\'enyi entropy for the aperiodic chain can be expressed as a sum of the entropy of the two possible blocks 
\be\label{eq37}
S_{q}(N)=\frac{\ln Y_{q}(N)}{1-q}=S_{q}^{a}(N_{a})+S_{q}^{b}(N_{b}),
\ee
where for each block, according to the eq.~\eqref{eq35} and ~\eqref{eq36}, the entropy corresponds to 
\be\label{eq38}
S_{q}^{a,b}(N_{a,b})=\sum_{i=1}^{N_{a,b}/2}\frac{\ln y_{q}(K_{a,b})}{1-q}+S_{q}^{R}(\frac{N_{a,b}}{2}).
\ee
With $S_{q}^{R}(\frac{N_{a,b}}{2})=\frac{\ln Y_{q}^{R}(\frac{N_{a,b}}{2})}{1-q}$, we denote the entropy of each renormalized block of the chain.

The Shannon-R\'enyi entropy of eq.~\eqref{eq38} as well as in the clean case of ref. \cite{month0} depends mainly on the factors $y_{q}^{a,b}(K_{a,b})$ of each block. The choice of block renormalization lead the 
Shannon-R\'enyi entropy in the first iteration to be
\be\label{eq_entr}
S_{q}(N)=\frac{N_a}{2}\frac{\ln y_{q}^{a}(K_{a})}{1-q} + \frac{N_b}{2}\frac{\ln y_{q}^{b}(K_{b})}{1-q}+ S_{q}^{R}.
\ee
The last term represents, the renormalized entropy. According to the BRG evolution which we discussed and presented in Fig.~\ref{fig:rg1}, the two control 
parameters $K_a$ and $K_b$ remain the same in each iteration $K_{a',b'}^{(j+1)}=K_{a,b}^{(j)}$, where with the index $j$ we denote the RG iteration number. The BRG rule for the Shannon-R\'enyi entropy can be written in the form
\be\label{eq:entro}
\begin{split}
S_{q}(N)& = \frac{N_a}{2}\frac{\ln y_{q}^{a}(K_{a}^{(j=1)})}{1-q} +  \frac{N_b}{2}\frac{\ln y_{q}^{b}(K_{b}^{(j=1)})}{1-q}\\
        & + \frac{N_a}{4}\frac{\ln y_{q}^{a}(K_{a}^{(j=2)})}{1-q} +  \frac{N_b}{4}\frac{\ln y_{q}^{b}(K_{b}^{(j=2)})}{1-q}+\cdot\cdot\cdot\\
        & + \frac{N_a}{2^k}\frac{\ln y_{q}^{a}(K_{a}^{(j=k)})}{1-q} +  \frac{N_b}{2^k}\frac{\ln y_{q}^{b}(K_{b}^{(j=k)})}{1-q},
\end{split}
\ee
\subsection{Generalized Multifractal Dimension}
From the eq.~\eqref{eq02} and following the analysis for the clean case of ref.\cite{month0} for the Shannon-R\'enyi entropy of eq. \eqref{eq_entr} and \eqref{eq:entro}, the generalized multifractal dimension $D_q(N)$ for the Period-Doubling aperiodic Ising chain, is given as sum of series of the RG of $k=1,2,..+\infty$ number of iterations on each block
\be\label{d1}
D_{q}(N)=\lim_{N\rightarrow \infty} \frac{S_{q}(N)}{N\ln2} = \sum_{k=1}^{+\infty}\frac{d_{q}(K_{a}) + d_{q}(K_{b})}{2^k},
\ee
with the auxiliary functions for each block
\be
d_{q}(K_{a})=N_{a}\frac{\ln y_{q}(K_{a})}{(1-q)\ln2}=\frac{2}{3}\frac{\ln\bigg(\bigg[\frac{1+\frac{K_{a}}{\sqrt{1+K_{a}^2}}}{2}\bigg]^{q}+\bigg[\frac{1-\frac{K_{a}}{\sqrt{1+K_{a}^2}}}{2}\bigg]^{q}\bigg)}{(1-q)\ln2}
\ee
and
\be
d_{q}(K_{b})=N_{b}\frac{\ln y_{q}(K_{b})}{(1-q)\ln2}=\frac{1}{3}\frac{\ln\bigg(\bigg[\frac{1+\frac{K_{b}}{\sqrt{1+K_{b}^2}}}{2}\bigg]^{q}+\bigg[\frac{1-\frac{K_{b}}{\sqrt{1+K_{b}^2}}}{2}\bigg]^{q}\bigg)}{(1-q)\ln2}.
\ee
Since we are interested for the dependence of the generalized fractal dimension $D_q$ on the coupling ratio $\rho=J_{1}/J_{0}$ for the quantum Ising chain with different type of aperiodic modulation, we repeat the BRG study of the subsections \ref{mul1} and ~\ref{mul2}, for the Thue-Morse (for more details see ~\ref{ap12}) and Rudin-Shapiro (for more details see ~\ref{ap22}) sequence of the table~\ref{tabl1}. For all the cases, we extract the $D_q$ similar to the eq.~\eqref{d1} and we perform a numerical simulation on $D_q$ for ten values of $\rho$ (to increase from $0$ to $1$). We locate our system on the critical point according to the eq. \eqref{eq:crp0} for the Period-Doubling and the corresponding functions for the other sequences.

\begin{figure}
    \centering
    \begin{subfigure}[b]{0.45\textwidth}
        \includegraphics[width=\textwidth]{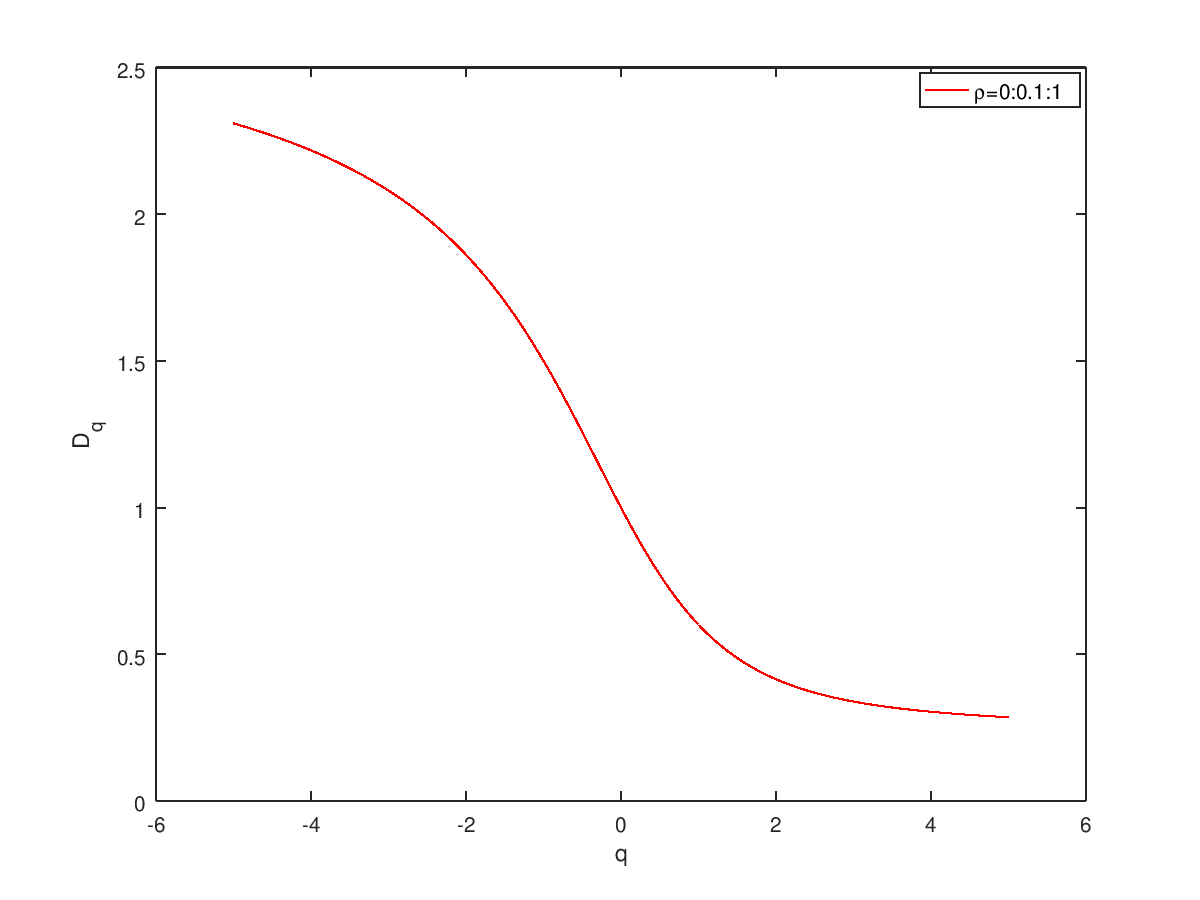}
        \caption{The Thue-Morse sequence}
        \label{fig:thu}
    \end{subfigure}
    ~ %add desired spacing between images, e. g. ~, \quad, \qquad, \hfill etc. 
    \begin{subfigure}[b]{0.45\textwidth}
        \includegraphics[width=\textwidth]{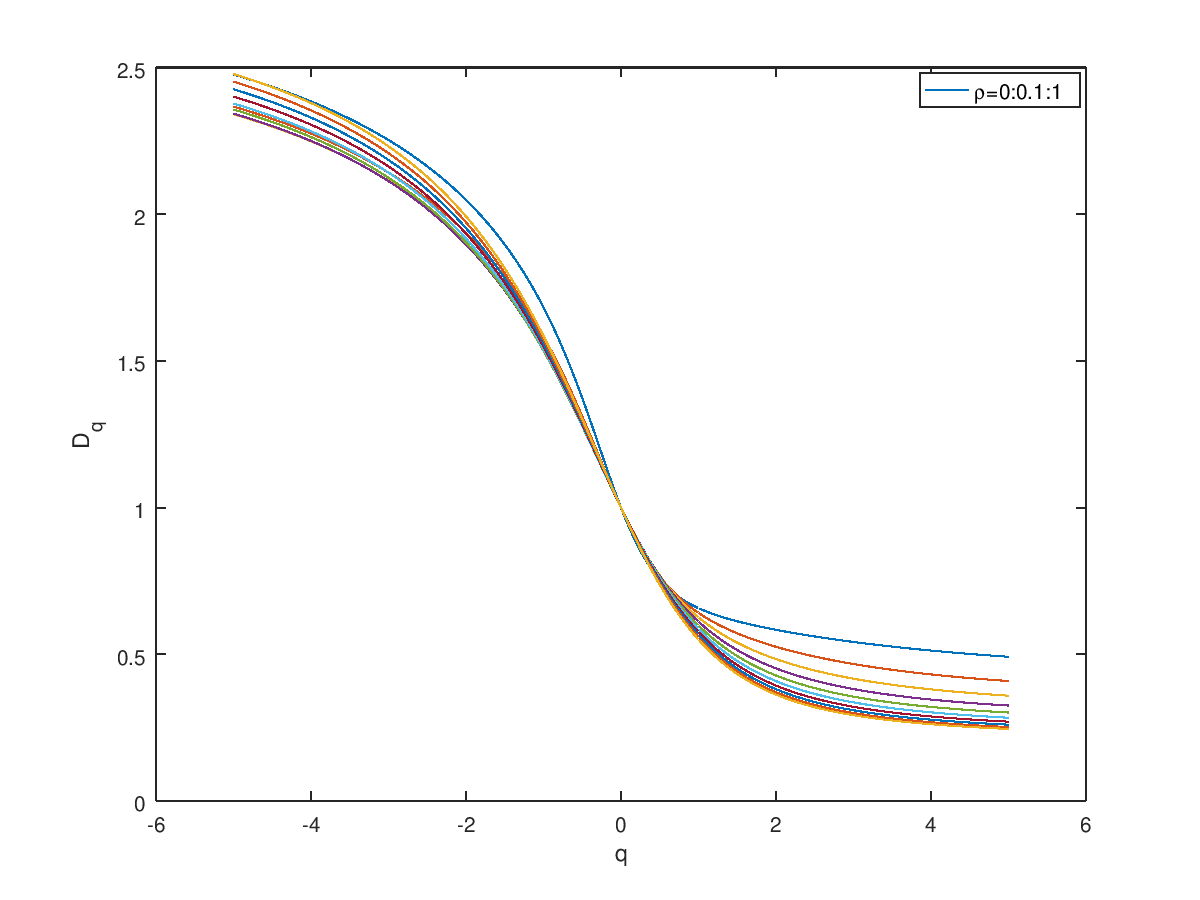}
        \caption{The Period-Doubling sequence}
        \label{fig:per}
    \end{subfigure}
    ~ %add desired spacing between images, e. g. ~, \quad, \qquad, \hfill etc. 
    \caption{Generalized Fractal Dimension $D_q$ as a function of R\'enyi index $q$ for ten values of the coupling ratio $\rho$.}\label{fig:thu-per}
\end{figure}

For the Thue-Morse sequence in Fig.~\ref{fig:thu}, with negative wandering exponent $\omega$, we observe that $D_q$ is independent of the coupling ratio $\rho$. The Fig.~\ref{fig:thu} is equal to the uniform quantum Ising chain \cite{atas1,month0}. On the other side, for the Period-Doublin sequence in Fig.~\ref{fig:per}, with vanishing wandering exponent, the $D_q$ vary with the coupling ratio. We observe that as the coupling ratio is increasing, the $D_q$ is slightly increasing for negative values of $q$ and decreasing for positive $q$. The dependence of $D_q$ with the coupling ratio for the marginal Period-Doubling sequence remind us the dependence of the dynamical exponent $z$ with the $\rho$ for the same sequence obtained with exact calculations \cite{igloi} as well as with Strong-Disorder RG approach \cite{vieira1,vieira2}. Both aperiodic sequences, present the usual non-linear behaviour which we discussed in the Introduction \ref{int}. 

\begin{figure}
  \centering
    \includegraphics[width=0.55\textwidth]{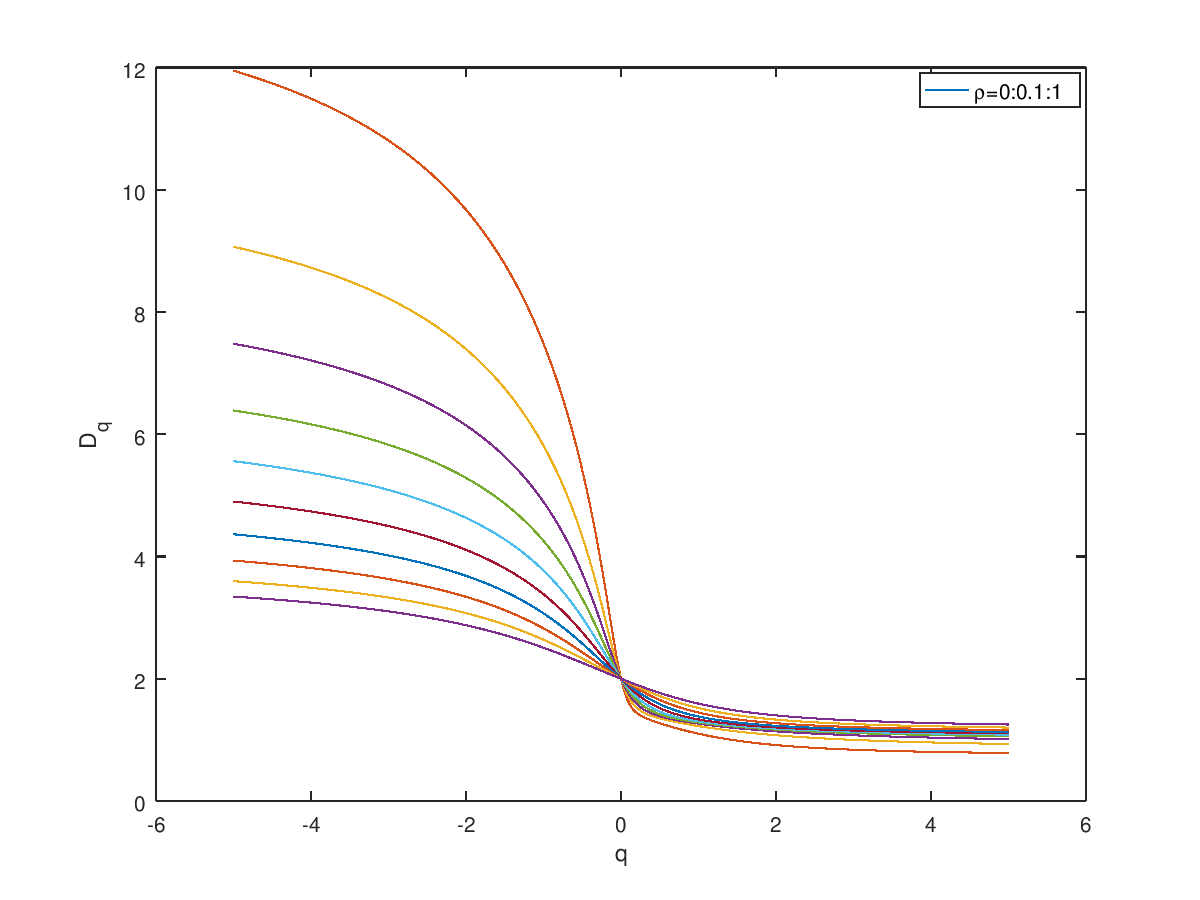}
     \caption{Generalized Fractal Dimension $D_q$ as a function of R\'enyi index $q$ for Rudin-Shapiro sequence.}
     \label{fig:rub}
\end{figure}

In the Fig.~\ref{fig:rub}, we present the $D_q$ for the Rudin-Shapiro aperiodic sequence with a positive wandering exponent $\omega$. We consider the case of the ref.~\cite{vieira1}, where they chose, $J_{b}=\rho J_a$, $J_{c}=\rho^2 J_a$ and $J_{d}=\rho^3 J_a$, for $0<\rho<1$. In this case, the $D_q$ has a different shape for negative values of the R\'enyi index $q$, in contrast to the previous aperiodic sequences. We observe that the spectrum of $D_q$ is larger for negative $q$ and smaller for positive. When the coupling ratio is equal to $\rho=1$, the diagram of $D_q$, is the same with the previous cases. For a values of $\rho\neq1$ the $D_q$ is widenly modified and is different from the Period-Doubling sequence of Fig.~\ref{fig:per}, where the coupling ratio was relevant in the diagram but was not able to lead to very different values of $D_q$. This can be understood by the fact that in Rudin-Shapiro sequence we have a bigger range of couplings and the inequality in the values can create a different spectrum of $D_q$. On the other side, the block renormalization approach \cite{month0,ferna-pach} that we are considering here, we believe that is able to capture the general characteristics of the $D_q$, while in cases with more complicated substitution rules, such Rudin-Shapiro sequence where multiple interaction couplings can create different kind of geometric fluctuations, a more careful block RG choice is required. 
\begin{table}
\begin{center}
    \begin{tabular}{| l | l | l | l | l | l |}
    \hline
    q & Exact Value & BRG \cite{month0} & TM & PD & RS \\ \hline
    1/2 & 0.841 \cite{atas1} & 0.771 & 0.760 & 0.723 & 1.720 \\ \hline
    1 & 0.610 \cite{step2} & 0.600 & 0.592 & 0.644 & 1.136 \\ \hline
    2 & 0.308 \cite{step2} & 0.415 & 0.424 & 0.474 & 0.897 \\
    \hline
    \end{tabular}
\end{center}
\caption{Special values of the $D_q$ for the Thue-Morse (TM), the Period-Doubling (PD) and  the Rudin-Shapiro (RS) sequences.}
\label{tab2}
 \end{table}
 
In the table ~\ref{tab2}, we compute the $D_q$ from the relation \eqref{d1}
for three different values of $q$, in order to compare them with the explicit exact values for $q=1/2$ of the ref. \cite{atas1,atas2} and the numerical values for $q=1$ and $q=2$ of the ref. \cite{step2}. We have to note that for the pure chain, the BRG results on the above values of $q$, they found to be close to the exact value for $q=1$ and far from the numerical values for $q\neq1$, as was presented in the ref. \cite{month0}. In our case, we considered the relation \eqref{eq:crp0} for the couplings on the critical point. For the Thue-Morse sequence, the values of $D_q$ are very close to the results of the ref. \cite{month0}. For the Period-Doubling sequence, the results are getting worst comparing the previous case. Finally, for the Rudin-Shapiro, the results of $D_q$ are very far from both the exact and the BRG results.

\subsection{Multifractal spectrum}
In order to calculate the multifractal spectrum $f(\a)$, we consider the saddle-point calculation in $\a$ for the IPR, where the multifractal spectrum can be parameterized in terms of the coefficients \cite{month0} as
\be\label{mf1}
\a_{q}=\frac{-\partial_{q}\ln Y_q}{\ln M} = -\frac{\sum_{m=1}^{M}\abs{\psi_{m}}^{2q}\ln \abs{\psi_{m}}^{2q}}{\sum_{m=1}^{M}\abs{\psi_{m}}^{2q} \ln M }
\ee
and
\be\label{mf2}
f(\a_{q})=-\frac{ \sum_{m=1}^{M}\abs{\psi_{m}}^{2q} \ln \frac{\abs{\psi_{m}}^{2q}}{\sum_{m=1}^{M}\abs{\psi_{m}}^{2q}}}{\sum_{m=1}^{M}\abs{\psi_{m}}^{2q} \ln M}
\ee
where $\a_q$ is the saddle-point value of $\a$ related to the exponent $\tau(q)$ by Legendre transform. 

According to the ref. \cite{month0} for the clean chain and considering the auxiliary functions ~\eqref{au1} and \eqref{au2} of IPR of the eq.~\eqref{eq2} and \eqref{eq3}, the equations for the multifractal spectrum can be written for each block as 
\be\label{w1}
\a_{q}(K_{a,b})=-\frac{1}{\ln2}\Bigg(\frac{\Big[1+\frac{K_{a,b}}{\sqrt{1+K_{a,b}^2}}\Big]^q \ln\Big[1+\frac{K_{a,b}}{\sqrt{1+K_{a,b}^2}}\Big] + \Big[1-\frac{K_{a,b}}{\sqrt{1+K_{a,b}^2}}\Big]^q \ln\Big[1-\frac{K_{a,b}}{\sqrt{1+K_{a,b}^2}}\Big]}{\Big[1+\frac{K_{a,b}}{\sqrt{1+K_{a,b}^2}}\Big]+\Big[1-\frac{K_{a,b}}{\sqrt{1+K_{a,b}^2}}\Big]} \Bigg)
\ee
and 
\be\label{w2}
f_{K_{a,b}}(\a_{q})=q\a_{q}(K_{a,b})+\frac{\ln\Big( \Big[1+\frac{K_{a,b}}{\sqrt{1+K_{a,b}^2}}\Big]^q +\Big[1-\frac{K_{a,b}}{\sqrt{1+K_{a,b}^2}}\Big]^q \Big)}{\ln2}.
\ee
Then, the BRG approach for the multifractal spectrum of the Period-Doubling sequence can be summarized as
\be\label{mf3}
\a_{q}=N_{a}\a_{q}(K_{a}) + N_{b}\a_{q}(K_{b})
\ee
and 
\be\label{mf4}
f(\a_{q})= N_{a} f_{K_a}(\a_q ) + N_b f_{K_b } (\a_q ),
\ee
where $N_a =2/3$ and $N_b =1/3$.

\begin{figure}
    \centering
    \begin{subfigure}[b]{0.45\textwidth}
        \includegraphics[width=\textwidth]{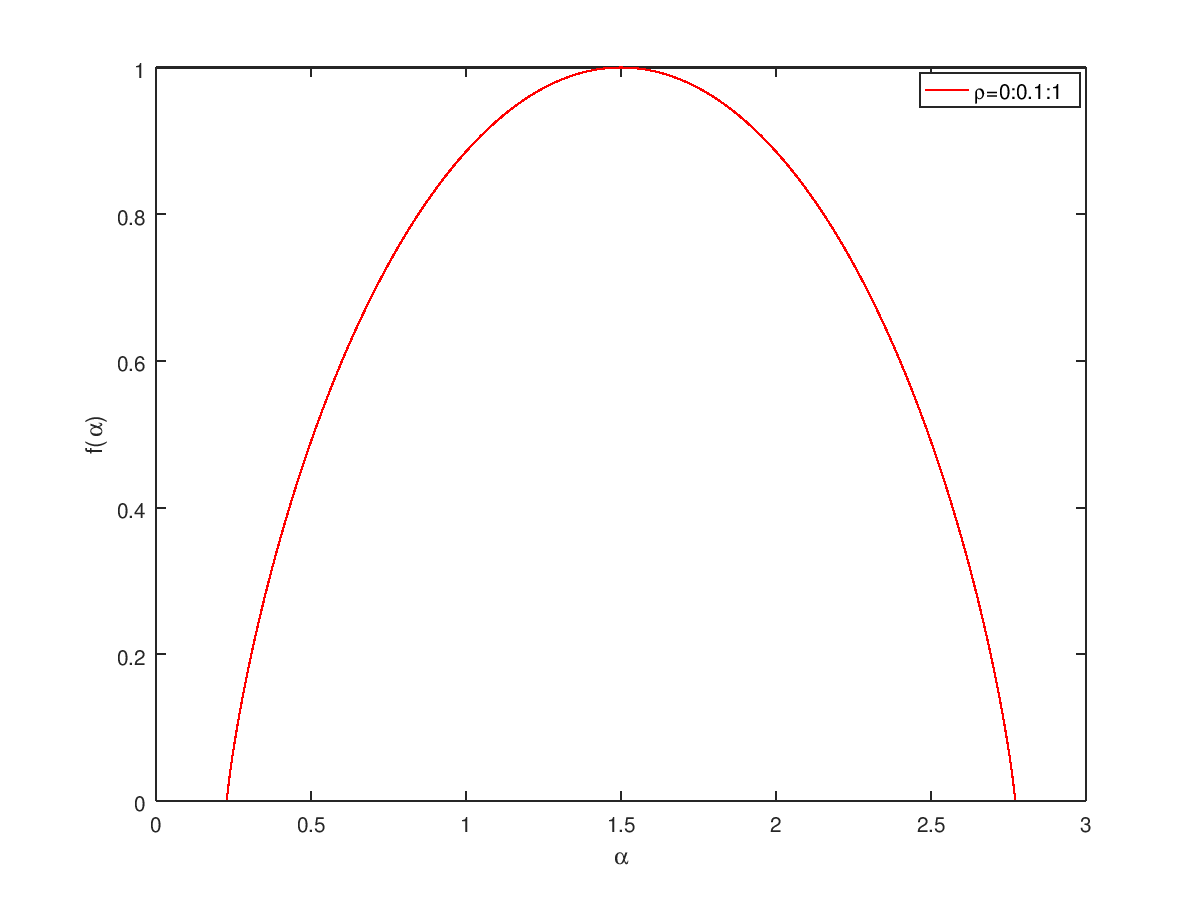}
        \caption{The Thue-Morse sequence.}
        \label{fig:fom}
    \end{subfigure}
    ~ %add desired spacing between images, e. g. ~, \quad, \qquad, \hfill etc. 
    \begin{subfigure}[b]{0.45\textwidth}
        \includegraphics[width=\textwidth]{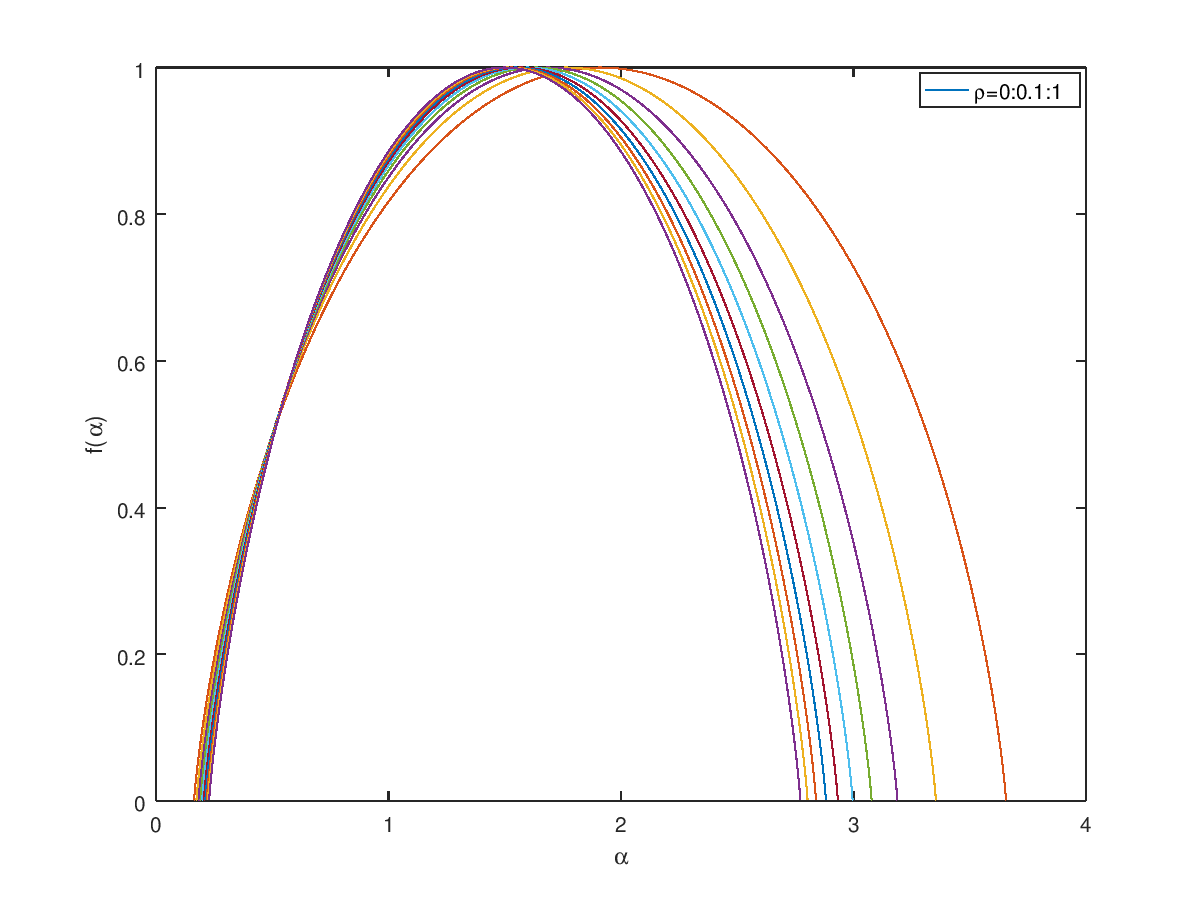}
        \caption{The Period-Doubling sequence.}
        \label{fig:fep}
    \end{subfigure}
    ~ %add desired spacing between images, e. g. ~, \quad, \qquad, \hfill etc. 
    \caption{Multifractal spectrum $f(\a)$ as function of $\a$ for the Thue-Morse and Period-Doubling sequences for different values of the coupling ratio $\rho$.}\label{fig:f1}
\end{figure}

We are interested in the dependence of the $f(\a)$ with the coupling ratio $\rho$ on the critical point of eq.~\eqref{eq:crp0}. In Fig.~\ref{fig:f1} we present the scaling of $f(\a)$ as extracted from the eq. \eqref{w1},\eqref{w2} and \eqref{mf3},\eqref{mf4} for the Thue-Morse and Period-Doubling sequences.
The first case of the Thue-Morse sequence, the coupling ratio is irrelevant for the multifractal spectrum. The Fig.~\ref{fig:fom} is equivalent with the clean chain \cite{atas2,month0}. On the other hand, for the Period-Doubling sequence, the coupling ratio $\rho$ is relevant for the $f(\a)$, as presented in the Fig.~\ref{fig:fep}. The spectrum $\a$ is found to be in the same region with the Thue-Morse case. The non-linear behavior of $f(\a)$ is the same with the clean case.
\begin{figure}
  \centering
    \includegraphics[width=0.55\textwidth]{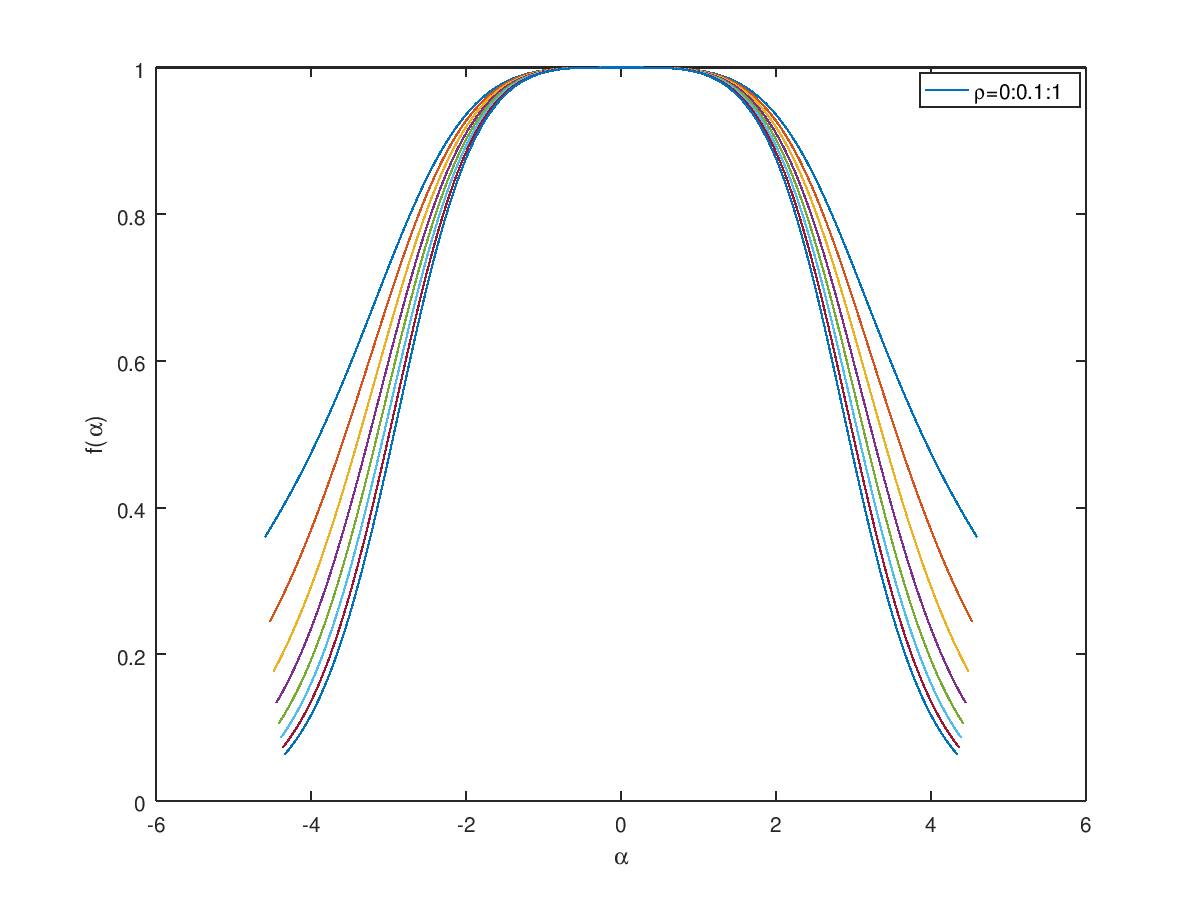}
     \caption{The multifractal spectrum $f(\a)$ as a function of $\a$ for different values of the coupling ration $\rho$ for the Rudin-Shapiro sequence.}
     \label{fig:rudf}
\end{figure}

Finally, the $f(\a)$ for the Rudin-Shapiro sequence is presented in Fig.~\ref{fig:rub}. The coupling ratio $\rho$ is relevant here and the spectrum $\a$ extends in larger region. The multifractal spectrum has similar form with the ref. \cite{luck2}, where the authors studied the multifractal spectrum of the Fourier spectrum of a general aperiodic sequence with positive wandering exponent. In our case, for the aperiodic quantum Ising chain, the multifractal spectrum has the same peak with the ref. \cite{luck2}, $f(\a)=1$ but for different value of the spectrum $\a=0$, instead $\a\approx1.29$. In our case, the spectrum takes negative and positive values since we consider the index $q$ in the region of $q=[-5,5]$. The authors of ref. \cite{luck2}, they consider only positive region of the index values where in our case, we are interested to compare with the results of the clean case of ref. \cite{atas1,month0}. 
%
%%%%%%%%%%%%%%%%%%%%%%%%%%%%%%%%%%%%%%%%%%%%%%%%%%%%%%%%%
%
\section{Conclusion}\label{end}
The context of multifractality investigated recently for the ground-state wavefunction of various quantum spin chains \cite{atas1,atas2} where it found to be multifractal in general and to have non-linear behavior with the index $q$. In this article, we studied the generalized fractal dimension $D_q$ and the multifractal spectrum $f(\a)$ for the aperiodic quantum Ising chain. We performed a block real-space renormalization procedure \cite{ferna-pach,month1} to obtain the ground-state wave functions and to extract the Infinite Participation Ratio and the related Shannon-R\'enyi entropy for three type of aperiodic perturbation on the critical point. From these quantities, we computed the $D_q$ and $f(\a)$.

For the irrelevant Thue-Morse sequence with negative wandering exponent, the system is independent of the coupling ratio $\rho$ and both quantities are similar to the unperturbed system. For the marginal Period-Doubling sequence with vanishing wandering exponent, the multifractal quantities vary with the coupling ratio but preserve the same non-linear behavior with previous case. Finally, for the relevant Rudin-Shapiro sequence with positive wandering exponent, we observe a different behavior for the $D_q$ and the different values of coupling ratio, dramatically affect the spectrum of $D_q$ and $f(\a)$.
We conclude that the quantum ground-state wavefunction of the aperiodic quantum Ising chain is multifractal. The resulting quantities, $D_q$ and $f(\a)$,  depends on the wandering exponent $\omega$. 
The summary follows the Luck criterion \cite{luck} in the sense that, the the wandering exponent control the behavior of the multifractality on the critical point. 

We have to note that our block real-space renormalization scheme, captures approximately the multifractal quantities as follows from the table \ref{tab2}, where we compare our results with the known exact values. In this case an exact diagonalization approach for the calculation of the ground-state wave function of the aperiodic chain is required and especially for the Rudin-Shapiro sequence. Our scope was to study the general characteristics of the evolution of $\omega$ to the multifractal quantities but an exact diagonalization approach will clarify the results in a better precision. Will be interesting to extend the BRG study of the multifractal quantities in other models with aperiodic perturbation, such as, Potts and Ashkin-Teller as well as to the three-site antiferromagnetic Ising model \cite{ref2}.
%...........................................
\section*{Acknowledgement}
We would like to thank C\'ecile Monthus and Christophe Chatelain for the useful discussions and comments in the initial stage of the project. Also, we would like to thank Nikos Fytas, Jos\'e A. Hoyos and Jo\~{a}o Getelina for the useful discussions, comments and collaboration on similar topics. Finally, we are grateful to Andr\'e P. Vieira, for the useful discussions and for pointing out the existence of the ref. \cite{luck2}. This study is supported from the Brazilian funding agency FAPESP under the thematic project of No: \emph{2017/11484-0}.
\appendix
\section{ The Thue-Morse sequence}
\subsection{Block RG rules and critical point}\label{ap11}
From the substitution rules, $0\rightarrow01$ and $1\rightarrow10$, we define the two blocks
\begin{equation*}
\begin{split}
& (\textbf{\textrm{block-a}}) \rightarrow 01\\
& (\textbf{\textrm{block-b}}) \rightarrow 10.
\end{split}
\end{equation*}
After the diagonalization of the two conventional Hamiltonians, we obtain the two lowest-lying energy states for each block
 \be
\begin{split}
 & \ket{\lambda_{i-1}^{-}(S_{0}^{a})}=\sqrt{\frac{1+\frac{J_{0}S_{0}^{a}}{\sqrt{J_{0}^{2}+h_{1}^{2}}}}{2}}\ket{S_{i-1}=+}+\sqrt{\frac{1-\frac{J_{0}S_{0}^{a}}{\sqrt{J_{0}^{2}+h_{1}^{2}}}}{2}}\ket{S_{i-1}=-}\\
 & \ket{\lambda_{i-1}^{-}(S_{0}^{b})}=\sqrt{\frac{1+\frac{J_{1}S_{0}^{a}}{\sqrt{J_{1}^{2}+h_{0}^{2}}}}{2}}\ket{S_{i-1}=+}+\sqrt{\frac{1-\frac{J_{1}S_{0}^{b}}{\sqrt{J_{1}^{2}+h_{0}^{2}}}}{2}}\ket{S_{i-1}=-}\\
\end{split}.
\ee
and the effective couplings and transverse fields
\be
J_{a}=\frac{J_{1}J_{0}}{\sqrt{J_{0}^{2}+h_{1}^2}}, \quad
h_{a}=\frac{h_{0}h_{1}}{\sqrt{J_{0}^{2}+h_{1}^2}}
\ee
\be
J_{b}=\frac{J_{0}J_{1}}{\sqrt{J_{1}^{2}+h_{0}^2}}, \quad
h_{b}=\frac{h_{1}h_{0}}{\sqrt{J_{1}^{2}+h_{0}^2}}.
\ee
From the effective couplings, we obtain the critical point
\be
h_{0}h_{1}=J_{0}J_{1}
\ee
\subsection{Generalized Multifractal Dimension}\label{ap12}
The generalized multifractal dimension, as we presented in eq.\eqref{d1} is given as a sum of the two auxiliary functions for each block
\be
d_{q}(K_{a})=N_{a}\frac{\ln y_{q}(K_{a})}{(1-q)\ln2}=\frac{1}{2}\frac{\ln\bigg(\bigg[\frac{1+\frac{K_{a}}{\sqrt{1+K_{a}^2}}}{2}\bigg]^{q}+\bigg[\frac{1-\frac{K_{a}}{\sqrt{1+K_{a}^2}}}{2}\bigg]^{q}\bigg)}{(1-q)\ln2},
\ee
\be
d_{q}(K_{b})=N_{b}\frac{\ln y_{q}(K_{a})}{(1-q)\ln2}=\frac{1}{2}\frac{\ln\bigg(\bigg[\frac{1+\frac{K_{b}}{\sqrt{1+K_{b}^2}}}{2}\bigg]^{q}+\bigg[\frac{1-\frac{K_{b}}{\sqrt{1+K_{b}^2}}}{2}\bigg]^{q}\bigg)}{(1-q)\ln2}
\ee
where $K_{a}=J_{0}/h_1$ and $K_{b}=J_{1}/h_0$, are the control parameters for each block.
\section{The Rudin-Shapiro sequence}\label{ap2}
\subsection{Block RG rules and critical point}\label{ap21}
From the substitution rules, $0\rightarrow01$, $1\rightarrow02$, $2\rightarrow31$ and $3\rightarrow32$, we define the following four blocks
\begin{equation*}
\begin{split}
& (\textbf{\textrm{block-a}}) \rightarrow 01\\
& (\textbf{\textrm{block-b}}) \rightarrow 02\\
& (\textbf{\textrm{block-c}}) \rightarrow 31\\
& (\textbf{\textrm{block-d}}) \rightarrow 32.
\end{split}
\end{equation*}
We consider four type of intra-part Hamiltonian for each block, we obtain the lowest-lying states for each block
 \be
\begin{split}
 & \ket{\lambda_{i-1}^{-}(S_{0}^{a})}=\sqrt{\frac{1+\frac{J_{0}S_{0}^{a}}{\sqrt{J_{0}^{2}+h_{1}^{2}}}}{2}}\ket{S_{i-1}=+}+\sqrt{\frac{1-\frac{J_{0}S_{0}^{a}}{\sqrt{J_{0}^{2}+h_{1}^{2}}}}{2}}\ket{S_{i-1}=-}\\
 & \ket{\lambda_{i-1}^{-}(S_{0}^{b})}=\sqrt{\frac{1+\frac{J_{0}S_{0}^{b}}{\sqrt{J_{0}^{2}+h_{2}^{2}}}}{2}}\ket{S_{i-1}=+}+\sqrt{\frac{1-\frac{J_{0}S_{0}^{b}}{\sqrt{J_{0}^{2}+h_{2}^{2}}}}{2}}\ket{S_{i-1}=-}\\
 & \ket{\lambda_{i-1}^{-}(S_{0}^{c})}=\sqrt{\frac{1+\frac{J_{3}S_{0}^{c}}{\sqrt{J_{3}^{2}+h_{1}^{2}}}}{2}}\ket{S_{i-1}=+}+\sqrt{\frac{1-\frac{J_{3}S_{0}^{c}}{\sqrt{J_{3}^{2}+h_{1}^{2}}}}{2}}\ket{S_{i-1}=-}\\
 & \ket{\lambda_{i-1}^{-}(S_{0}^{d})}=\sqrt{\frac{1+\frac{J_{3}S_{0}^{d}}{\sqrt{J_{3}^{2}+h_{2}^{2}}}}{2}}\ket{S_{i-1}=+}+\sqrt{\frac{1-\frac{J_{3}S_{0}^{d}}{\sqrt{J_{3}^{2}+h_{2}^{2}}}}{2}}\ket{S_{i-1}=-}.\\
\end{split}.
\ee
The effective couplings and transverse fields are
\be
J_{a}=\frac{J_{1}J_{0}}{\sqrt{J_{0}^{2}+h_{1}^2}}, \quad
h_{a}=\frac{h_{0}h_{1}}{\sqrt{J_{0}^{2}+h_{1}^2}}
\ee
\be
J_{b}=\frac{J_{2}J_{0}}{\sqrt{J_{0}^{2}+h_{2}^2}}, \quad
h_{b}=\frac{h_{0}h_{2}}{\sqrt{J_{0}^{2}+h_{2}^2}}.
\ee
\be
J_{c}=\frac{J_{1}J_{3}}{\sqrt{J_{3}^{2}+h_{1}^2}}, \quad
h_{c}=\frac{h_{1}h_{3}}{\sqrt{J_{3}^{2}+h_{1}^2}}.
\ee
\be
J_{d}=\frac{J_{2}J_{3}}{\sqrt{J_{3}^{2}+h_{2}^2}}, \quad
h_{b}=\frac{h_{3}h_{2}}{\sqrt{J_{3}^{2}+h_{2}^2}}.
\ee
Since in this case, they involved four blocks, we consider the asymptotic density of each block, $\rho_{\infty}^{a}=\rho_{\infty}^{b}=\rho_{\infty}^{c}=\rho_{\infty}^{d}=1/4$. Therefore, from our analysis on eq.~\eqref{cri_poi}, the critical point for the Rudin-Shapiro sequence is given by the relation
\be
h_{0}h_{1}h_{2}h_{3}=J_{0}J_{1}J_{2}J_{3}.
\ee
\subsection{Generalized Multifractal Dimension}\label{ap22}
The generalized multifractal dimension, as we presented in eq.\eqref{d1} is given as a sum of the four auxiliary functions for each block
\be\label{ape1}
d_{q}(K_{\mu})=N_{\mu}\frac{\ln y_{q}(K_{\mu})}{(1-q)\ln2}=\frac{1}{2}\frac{\ln\bigg(\bigg[\frac{1+\frac{K_{\mu}}{\sqrt{1+K_{\mu}^2}}}{2}\bigg]^{q}+\bigg[\frac{1-\frac{K_{\mu}}{\sqrt{1+K_{\mu}^2}}}{2}\bigg]^{q}\bigg)}{(1-q)\ln2},
\ee
where for simplicity we refer as $\mu=a,b,c,d$, since each block occupies the same part of the sequence. The control parameters are given as a functions of couplings and fields as, $K_{a}=J_{0}/h_{1}$, $K_{b}=J_{0}/h_{2}$, $K_{c}=J_{3}/h_{1}$ and $K_{d}=J_{3}/h_{2}$. Then, the generalized fractal dimension is given as a function of the auxiliary function of eq.~\eqref{ape1} for each block.
%..........................................................
%               Bibliography
%.........................................................

\medskip
 \section*{References}
\bibliographystyle{unsrt}

%\bibliography{ela}

\begin{thebibliography}{}

\end{thebibliography}


\begin{thebibliography}{99}
%
\bibitem{sach}
 S. Suchdev, Quantum Phase Transitions, Cambridge University Press; 2 edition (May 9, 2011).
%
\bibitem{kada}
 T.C. Halsey, M.H. Jensen, L.P. Kadanoff, I. Procaccia, B.I. Shraiman, \textit{Phys. Rev. A}\href{https://journals.aps.org/pra/abstract/10.1103/PhysRevA.33.1141}{\textbf{33}, 1141 (1986)}.
 %
 \bibitem{an1}
 F. Wegner, \textit{Z. Phys. B}\href{https://link.springer.com/article/10.1007/BF01325284}{\textbf{36}, 209 (1980)}.
 %
 \bibitem{an2}
 C. Castellani and L. Peliti, \textit{J. Phys. A}\href{https://iopscience.iop.org/article/10.1088/0305-4470/19/17/009}{ \textbf{19}, L429 (1986)}.
 %
 \bibitem{mirlin1}
 A. D. Mirlin and F. Evers, \textit{Phys. Rev. B}\href{https://journals.aps.org/prb/abstract/10.1103/PhysRevB.62.7920}{\textbf{62}, 7920 (2000)}.
 %
\bibitem{ref1}
B. B. Mandelbrot, \textit{The Fractal Geometry of Nature}, Freeman, New York, (1982).
%
\bibitem{ak}
L. de Arcangelis, S. Redner, and A. Coniglio, \textit{Phys. Rev. } \href{https://journals.aps.org/prb/abstract/10.1103/PhysRevB.31.4725}{\textbf{B 31}, 4725(R) (1985)}.
%
%\bibitem{procha}
%I. Procaccia, \textit{J. Stat. Phys.} \href{https://link.springer.com/article/10.1007/BF01012929}{ \textbf{36}, 649 (1984)}.
%
%
 %\bibitem{kada1}
%L.P. Kadanoff, \textit{Physics.} \href{https://journals.aps.org/ppf/abstract/10.1103/PhysicsPhysiqueFizika.2.263}{ \textbf{2}, 263 (1966)}.
 %
 
 %
 \bibitem{che}
 H. E. Stanley and P. Meakin, \textit{Nature}\href{https://www.nature.com/articles/335405a0} { \textbf{335}, 29 (1988)}.
 %
 \bibitem{you}
 T. Olsson and A.P. Young, \textit{Phys. Rev. B}\href{https://journals.aps.org/prb/abstract/10.1103/PhysRevB.60.3428} {\textbf{60}, 3428 (1999)}.
 %
 \bibitem{chr}
 C. Chatelain and B. Berche, \textit{Nucl. Phys., B }\href{https://www.sciencedirect.com/science/article/pii/S055032130000050X?via%3Dihub} {\textbf{572}, 626 (2000)}.
 %
 \bibitem{eve}
 F. Evers and A. D. Mirlin, \textit{Rev. Mod. Phys.} \href{https://journals.aps.org/rmp/abstract/10.1103/RevModPhys.80.1355} {\textbf{80}, 1355 (2008)}
 %
 \bibitem{exp1}
  A.  Richardella,  P.  Roushan,  S.  Mack,  B.  Zhou,  D.  A.  Huse, D. D. Awschalom and A. Yazdani, \textit{Science} \href{https://science.sciencemag.org/content/327/5966/665}{\textbf{327}, 665 (2010)}.
  %
  \bibitem{exp2}
   J. Chab\'e, G. Lemari\'{e}, B. Gremaud, D. Delande, P. Szriftgiserand J.-C. Garreau, \textit{Phys. Rev. Lett.} \href{https://journals.aps.org/prl/abstract/10.1103/PhysRevLett.101.255702} {\textbf{101}, 255702 (2008)}.
   %
   \bibitem{exp3}
   G. Lemarié, H. Lignier, D. Delande, P. Szriftgiser, and J.-C. Garreau, \textit{Phys. Rev. Lett.} \href{https://journals.aps.org/prl/abstract/10.1103/PhysRevLett.105.090601} {\textbf{105}, 090601 (2010).} 
 %
 \bibitem{atas1}
 Y.Y. Atas and E. Bogomolny, \textit{Phys. Rev. E}\href{https://journals.aps.org/pre/abstract/10.1103/PhysRevE.86.021104} {\textbf{86}, 021104 (2012)}.
 %
 \bibitem{atas2}
 Y.Y. Atas and E. Bogomolny, \textit{Phil. Trans. R. Soc. \textbf{A 372}}, 20120520 (2014).
 %
 \bibitem{lieb}
 E. Lieb, T. Schultz and D. Mattis, \textit{Ann. Phys.} \href{https://www.sciencedirect.com/science/article/pii/0003491661901154} {
\textbf{16} 407 (1961)}.
%
\bibitem{step1}
J.-M. Stéphan G. Misguich, and V. Pasquier, \textit{Phys. Rev. B} \href{https://journals.aps.org/prb/abstract/10.1103/PhysRevB.80.184421} {\textbf{80}, 184421 (2009)}.
%
\bibitem{step2}
J.-M. Stéphan, G. Misguich, and V. Pasquier, \textit{Phys. Rev. B} \href{https://journals.aps.org/prb/abstract/10.1103/PhysRevB.82.125455} {
\textbf{82}, 125455 (2010); ibid, 84, 195128 (2011)}.
%
\bibitem{month0}
C. Monthus, \textit{J. Stat. Mech.} \href{https://iopscience.iop.org/article/10.1088/1742-5468/2015/04/P04007/meta} {P04007 (2015)}.
%
%
\bibitem{month1}
C. Monthus, \textit{J. Stat. Mech.}\href{https://iopscience.iop.org/article/10.1088/1742-5468/2015/01/P01023/meta} { P01023 (2015)}.
%
\bibitem{quasi}
D. Schechtman, I. Blech, D. Gratias and J.W. Cahn, \textit{Phys. Rev. Lett.} \href{https://journals.aps.org/prl/abstract/10.1103/PhysRevLett.53.1951} { \textbf{53} 1951 (1984)}.
%
\bibitem{boo}
J. M. Dumont, in \textit{Number Theory and Physics, Spring. Proc. Physics}, Vol.\textbf{47}, edited by J. M.Luck, P. MoussaandM. Waldschmidt, (Springer, Berlin) (1990).
%
\bibitem{harris}
A.B. Harris, \textit{J. Phys. C: Solid State Phys.}\href{https://iopscience.iop.org/article/10.1088/0022-3719/7/9/009/pdf} {
\textbf{7} 1671 (1974)}.
%
\bibitem{luck}
J.M. Luck, \textit{Europhys. Lett.} \href{https://iopscience.iop.org/article/10.1209/0295-5075/24/5/007} {\textbf{24} 359 (1993)}.
%
\bibitem{loic}
L. Turban, F. Igloi and B. Berche, \textit{ Phys. Rev. B} \href{https://journals.aps.org/prb/abstract/10.1103/PhysRevB.49.12695} {\textbf{49},12695 (1994)}. 
%
\bibitem{herm}
J. Hermisson, U. Grimm and M. Baake M, \textit{J. Phys. A: Math. Gen.}  \href{https://iopscience.iop.org/article/10.1088/0305-4470/30/21/009/meta} {\textbf{30}, 7315 (1997)}.
%
\bibitem{ferna-pach}
A. Fernandez-Pacheco, \textit{Phys. Rev. D}\href{https://journals.aps.org/prd/abstract/10.1103/PhysRevD.19.3173} {\textbf{19}, 3173 (1979)}.
%
\bibitem{pot}
J. Solyom and P. Pfeuty, \textit{Phys. Rev. \textbf{B 24}}, 218 (1981);
J. Solyom, \textit{Phys. Rev. \textbf{B 24}}, 230 (1981).
%
\bibitem{ash}
F. Igloi and J. Solyom, J. Phys. \textit{C Solid State Physics \textbf{16}}, 2833 (1983);
F. Igloi and J. Solyom, \textit{Phys. Rev. B}\href{https://journals.aps.org/prb/abstract/10.1103/PhysRevB.28.2785}{\textbf{28}, 2785 (1983)};
F. Igloi and J. Solyom, \textit{J. Phys. A Math. Gen.} \href{https://iopscience.iop.org/article/10.1088/0305-4470/17/7/021/meta}{\textbf{17}, 1531 (1984)}.
%
\bibitem{nish1}
R. Miyazaki, H. Nishimori and G. Ortiz, \textit{Phys. Rev. E} \href{https://journals.aps.org/pre/abstract/10.1103/PhysRevE.83.051103} {\textbf{83}, 051103 (2011)}.
%
\bibitem{nish2}
R. Miyazaki and H. Nishimori, \textit{Phys. Rev. E}\href{https://journals.aps.org/pre/abstract/10.1103/PhysRevE.87.032154} {\textbf{87}, 032154 (2013)}.
%
\bibitem{fisher}
D. S. Fisher, \textit{Phys. Rev. Lett.}\href{https://journals.aps.org/prl/abstract/10.1103/PhysRevLett.69.534} {\textbf{69}, 534 (1992)}; \textit{Phys. Rev. B} \href{https://journals.aps.org/prb/abstract/10.1103/PhysRevB.51.6411}{\textbf{51}, 6411 (1995)}.
%
\bibitem{month2}
C. Monthus, \textit{J. Stat. Mech.} P033101 (2016).
% 
\bibitem{pfe}
P. Pfeuty, \textit{Ann. Phys. 57,} \href{https://www.sciencedirect.com/science/article/pii/0003491670902708} {79 (1970)}.
%
\bibitem{igloi}
F. Igl\'oi, L. Turban, D. Karevski, and F. Szalma, \textit{Phys. Rev. B} \href{https://journals.aps.org/prb/abstract/10.1103/PhysRevB.56.11031} {\textbf{56}, 11031 (1997)}.
%
\bibitem{vieira1}
F. J. Oliveira Filho, M. S. Faria,  and A. P. Vieira, \textit{J. Stat. Mech.}
\href{https://iopscience.iop.org/article/10.1088/1742-5468/2012/03/P03007/meta} {P03007 (2012)}.
%
\bibitem{vieira2}
A.P. Vieira, \textit{Phys. Rev. Lett.} \href{https://journals.aps.org/prl/abstract/10.1103/PhysRevLett.94.077201} {\textbf{94} 077201 (2005)}; \textit{Phys. Rev. B} \href{https://journals.aps.org/prb/abstract/10.1103/PhysRevB.71.134408} { \textbf{71}, 134408 (2005)}.
%
\bibitem{luck2}
C. Gord\'ece and J.M. Luck, \textit{Phys. Rev. B}\href{https://journals.aps.org/prb/abstract/10.1103/PhysRevB.45.176} {\textbf{45}, 176 (1992)}.
%
\bibitem{ref2}
N. S. Ananikian and S. K. Dallakian,
\textit{Physica D} \href{https://www.sciencedirect.com/science/article/pii/S0167278997000602}{107, 75 (1997)}.
\end{thebibliography}
\end{document}